\documentclass[aps,pra,superscriptaddress,twocolumn,a4paper,reprint,eprint,longbibliography, footinbib, floatfix]{revtex4-2}
\pdfoutput=1
\usepackage[percent]{overpic}
\usepackage{physics}
\usepackage[utf8]{inputenc}
\usepackage{graphicx}
\usepackage{amsmath}
\usepackage{amsfonts}
\usepackage{amssymb}
\usepackage{bbm}
\usepackage{bm}
\usepackage{bbold}
\usepackage[usenames,dvipsnames,svgnames,table]{xcolor}
\usepackage[USenglish]{babel}
\usepackage{hyperref}
\usepackage{grffile}
\usepackage{multirow}
\usepackage{hhline}

\newcommand{\clr}{\color{red}}

\newcommand{\lamp}{\multicolumn{1}{>{\columncolor{blue!25}[2pt]}c|}{$+1$}}
\newcommand{\lamn}{\multicolumn{1}{>{\columncolor{red!25}[2pt]}c|}{$-1$}}

\newcommand{\eq}[1]{Eq.~(\ref{#1})} 
\newcommand{\ignore}[1]{}

\allowdisplaybreaks

 \pdfpageattr{/Group <</S /Transparency /I true /CS /DeviceRGB>>}

\begin{document}

\title{General theory of robustness against disorder in multi-band superconductors}

\author{D. C. Cavanagh}
\email{david.cavanagh@otago.ac.nz}
\affiliation{Department of Physics, University of Otago, P.O. Box
56, Dunedin 9054, New Zealand}
\author{P. M. R. Brydon}
\email{philip.brydon@otago.ac.nz}
\affiliation{Department of Physics and MacDiarmid Institute for
Advanced Materials and Nanotechnology, University of Otago, P.O. Box
56, Dunedin 9054, New Zealand}

\date{14 April, 2021}

\begin{abstract}
We investigate the influence of general forms of disorder on the
robustness of superconductivity in
multiband materials. Specifically, we consider a general
  two-band system where the bands arise from an orbital degree of
  freedom of the electrons. Within the Born approximation, we show that the interplay of the
  spin-orbital structure of 
  the normal-state Hamiltonian, disorder scattering, and
  superconducting pairing potentials can lead to significant
  deviations from the expected robustness of the
  superconductivity. This can be conveniently formulated in terms of
  the so-called ``superconducting fitness''. In particular, we verify a key role for unconventional $s$-wave
  states, permitted by the spin-orbital structure and which may pair electrons
that are not time-reversed partners. To exemplify the role of Fermi surface topology and spin-orbital
  texture, we apply our formalism to the candidate topological superconductor
Cu$_x$Bi$_2$Se$_3$, for which only a single band crosses the Fermi
energy, as well as models of the iron pnictides, which possess
multiple Fermi pockets.
\end{abstract}
\maketitle
\section{Introduction}

The influence of disorder on superconductivity is a significant, and
frequently utilized, probe of the superconducting order parameter
\cite{Mineev1999}. When pairing occurs between time-reversed partners,
Anderson's theorem \cite{Anderson1959} states that disorder can depair
electrons only when the superconducting gap is anisotropic or the
disorder breaks time-reversal symmetry. The behavior of the critical
temperature $T_c$ in the presence of disorder is, as a result, one of
the key indicators of unconventional superconductivity in single band
materials. Conventional (fully gapped) superconductivity is only
  sensitive to time-reversal symmetry breaking (TRSB), or magnetic,
  disorder, while unconventional superconductors are equally
  susceptible to both time-reversal symmetric (TRS) disorder and TRSB
  disorder.

In many superconductors of recent interest, the low-energy electronic
states are conveniently labelled by discrete quantum numbers
additional to spin, e.g. the atomic orbital or sublattice site from
each the electrons originate. 
The existence of these novel 
``orbital'' degree of freedom has been proposed to play an important role in both the
normal state and superconducting properties \cite{Qi2011, Nomoto2016b, BlackSchaffer2013, Ramires2016, Yi2017, Savary2017}.
In such superconductors, pairing may be isotropic in momentum without
pairing time-reversed states \cite{FuBerg2010, Vafek2017,
  Agterberg2017FeSe, Brydon2016, Kawakami2018, Moeckli2018, Oiwa2018,
  Ong2016, Nica2017}, and there has hence been much recent interest in generalizing Anderson's theorem to account for such systems \cite{MichaeliFu2012, Cavanagh2020, Scheurer2015, Scheurer2016, ScheurerTimmons2020, Ramires2019, Sato2020, Dentelski2020}.
The effect of disorder in such systems is
considerably more complicated due to the interplay of the 
 internal spin-orbital structure of the
superconducting states with the spin-orbital texture
of the electronic system.  

Previously, we developed a framework to illustrate the significant role played by the spin-orbital texture in determining the robustness of various superconducting states, even those with momentum-dependent pairing functions, against scalar TRS disorder with no dependence on the internal degrees of freedom \cite{Cavanagh2020}. In this work, we extend this framework to consider both TRS and TRSB disorder with a non-trivial dependence on the internal degrees of freedom. Additionally, we highlight the increased robustness of anisotropic pairing states due to the existence of the unconventional $s$-wave states, and demonstrate how the number of bands at the Fermi level influences the robustness.

As a demonstration of the utility of our framework, we apply it to two families of materials proposed to realize unconventional $s$-wave states: superconducting Dirac systems (such as the apparently nematic and fully-gapped superconductor Cu$_x$Bi$_2$Se$_3$ \cite{FuBerg2010, Yip2013, Matano2016, Yonezawa2016, Tao2018, Fu2014}) and the iron pnictide superconductors \cite{Cvetkovic2013, Vafek2017}.  
The orbital degree of freedom has a significant influence on the superconductivity in the Dirac materials,  though only a single band crosses the Fermi level \cite{FuBerg2010, Yip2013, Fu2014, MichaeliFu2012,Dentelski2020, Sato2020, Ramires2019}. 
We demonstrate, for Cu$_x$Bi$_2$Se$_3$, the key role played by the spin-orbital structure of the disorder potential in determining the robustness of various superconducting states.

%

In the iron-based pnictide superconductors, multiple bands cross the
Fermi level leading to a Fermi surface with multiple sheets. The
consensus view of experiment and theory is that an $s^\pm$-wave state
is realized in the majority of these materials, where each Fermi
sheet has a largely isotropic gap but with opposite sign between the
electron- and hole-like Fermi surfaces~\cite{Mazin2008, Chubukov2008,
  Chubukov2012, Paglione2010}.
There is an extensive literature on the effect of impurities on this
pairing state, see e.g.~\cite{Onari2009,Efremov2011,Efremov_2013,Yamakawa2013,Kreisel2013,Stanev2014,Hoyer2015}.
Nevertheless, a variety of more exotic superconducting states have been
proposed that exploit the striking orbital texture of the Fermi surfaces~\cite{Daghofer2010,Ong2016,Vafek2017}. 
 Motivated by these works, and to emphasize the important role of
  Fermi surface spin-orbital polarization and topology, we examine a
  number of superconducting states which are possible 
in a well-known two-orbital model~\cite{Raghu2008} for these
materials.

Our paper is organized as follows: In Sec. \ref{Sec:Theory}, we develop the generalization of our theoretical framework, based on the self-consistent Born approximation, to account for the influence of non-scalar disorder on superconductivity in two-band systems. In Sec. \ref{Sec:BiSe}, we apply this framework to models of superconducting Dirac materials and highlight that the time-reversal symmetry, or lack thereof, of the disorder scattering is not the dominant factor responsible for determining the robustness of a given superconducting state. We consider, in Sec. \ref{Sec:FeAs}, the additional influence of multiple Fermi surface pockets in more detail by applying our framework to the iron pnictide superconductors. 
In Sec. \ref{Sec:Discussion}, we discuss some general insights, in particular the important role played by the superconducting fitness \cite{Ramires2016, Ramires2018, Cavanagh2020}, and make reconcile discrepancies between some recent results and our own.

\section{Theory}\label{Sec:Theory}

Our general framework is developed for a generic two-band system with both inversion and time-reversal symmetry. States in such a system are typically defined by the electron spin and a quantum number associated with electron orbital, sublattice or some other additional degree of freedom. In general, we refer to the four degrees of freedom as a `spin-orbital' basis, regardless of the origin of the additional degree of freedom. 

\subsection{Normal state properties}

We consider systems with the normal-state Hamiltonian  $H = \sum_{\bm
  k}c^\dagger_{\bm k}{\cal H}_{\bf k}c_{\bm k}$, where $c_{\bm k}$ is
a four component spinor encoding the internal degrees of
freedom. The most general form of the matrix  ${\cal H}_{\bm k}$ is~\cite{Brydon2018b},
\begin{equation}
 {\cal H}_{\bm k} = \epsilon_{{\bm k},0}\,\mathbb{1}_4 + \vec{\epsilon}_{\bm k}\cdot\vec{\gamma}\,, \label{eq:genH}
\end{equation}
where $\mathbb{1}_4$ is the $4\times4$ unit matrix and
$\vec{\gamma}=(\gamma^1,\gamma^2,\gamma^3,\gamma^4,\gamma^5)$ are the 
five mutually anti-commuting Euclidean Dirac matrices. The
coefficients of these matrices $\epsilon_{{\bm k},0}$ and
$\vec{\epsilon}_{\bm k} =(\epsilon_{{\bm k},1},\epsilon_{{\bm
    k},2},\epsilon_{{\bm k},3},  \epsilon_{{\bm k},4},\epsilon_{{\bm
    k},5})$ are all real functions. The eigenvalues
  of~\eq{eq:genH} are the band energies 
\begin{equation}
E_{{\bm k},\pm}=\epsilon_{{\bm k},0} \pm \left|\vec{\epsilon}_{\bm k}\right|.
\end{equation}
 The anticommutation of the $\gamma$ matrices among themselves
  ensures that these 
  eigenvalues are doubly degenerate, reflecting the presence of
  inversion and time-reversal symmetry. 
Time-reversal is given by the operator ${\cal T}=U_T{\cal K}$,
  where ${\cal K}$ is complex conjugation and the unitary part of the
  time-reversal operator can be chosen as $U_T=\gamma^3\gamma^5$
  without loss of generality.  Inversion symmetry either affects the
  internal degrees of freedom trivially (${\cal I}=\mathbb{1}_4$) or
  nontrivially (${\cal I}=\gamma^1$).  Due to the presence of
    time-reversal and inversion symmetry, it is generally
possible to label the two-fold-degenerate eigenstates of the
Hamiltonian in terms of a pseudospin index, which behaves like a
spin-$\frac{1}{2}$ under these two symmetries. Several authors have
analyzed the impurity problem in terms a band-pseudospin or similar
basis~\cite{Fu2014,Dentelski2020}. Although this has the advantage of
casting the pairing potentials in a more familiar form, we do not
pursue this approach as it obscures the important role of the
orbital-spin degree of freedom.

We consider isotropic scattering off potential impurities of
  different types $\alpha$ distributed randomly at positions ${\bm r}_{j_\alpha}$,  described by the Hamiltonian
\begin{equation}
  H_{\text{imp}} =
  \frac{1}{\Omega}\sum_{\alpha}\sum_{j_\alpha}\sum_{{\bm k},{\bm
      k}'}e^{i({\bm k}'-{\bm k})\cdot{\bm
      r}_{j_\alpha}}c^{\dagger}_{{\bm k}}\tilde{V}_\alpha c_{{\bm k}'} \label{eq:impHam}
\end{equation}
where $\Omega$ is the volume and $\tilde{V}_\alpha=V_\alpha\gamma^{\alpha_x}\gamma^{\alpha_y}$ is
the impurity potential. The choice $\alpha_x=\alpha_y=0$
 corresponds to the scalar disorder considered in
Ref. \cite{Cavanagh2020}. Note that our theory allows both for 
distinct impurities of different types  (i.e. ${\bf r}_{j_\alpha}\neq
{\bf r}_{j_{\tilde{\alpha}}}$ for $\alpha \neq \tilde{\alpha}$), or for the impurities to have
multiple different scattering potentials (i.e. ${\bf r}_{j_\alpha}=
{\bf r}_{j_{\tilde{\alpha}}}$). 
Within the Born approximation, the effect of the impurities is
accounted for via a self-energy 
$\Sigma_1(i\omega_n)$, so that the full Green's function satisfies the
Dyson equation
\begin{equation}
\bar{G}^{-1}({\bm k},i\omega_n) = G_0^{-1}({\bm k},i\omega_n) -
\Sigma_1(i\omega_n)
\end{equation}
where $G_0({\bm k},i\omega_n)$ is the Green's function of the clean
system and the self-energy is determined self-consistently 
\begin{equation}
\Sigma_{1}(i\omega_n) = \sum_{\alpha}n_{\text{imp},\alpha}\tilde{V}_\alpha\int\!\!\frac{d^3k}{(2\pi)^3}\bar{G}({\bm 	k},i\omega_n)\tilde{V}^\dagger_\alpha\,,
\end{equation}
where $n_{\text{imp},\alpha}$ is the concentration of $\alpha$-type impurities. 
This approximation is valid when disorder scattering is weak
relative to the chemical potential, $\hbar \tau^{-1}\ll \mu$, where
$\tau^{-1}$ is the disorder scattering rate. If we further require
that the disorder scattering is small compared to the band separation
$|\vec{\epsilon}_{\bm k}|$ at the Fermi surface, 
then to leading order in $\hbar \tau^{-1}/|\vec{\epsilon}_{\bm k}|$  the  Green's
functions of the disordered system is 
\begin{equation}
      \bar{G}({\bm k},i\omega_n) = \sum_{j=\pm}\frac{1}{i\tilde{\omega}_{n,j} - E_{{\bm k},j}}{\cal P}_{{\bm k},j} \label{eq:GF}
\end{equation}
where ${\cal P}_{{\bm k},\pm} = \frac{1}{2}\left(\mathbb{1}_4
  \pm\hat{\epsilon}_{\bm k}\cdot\vec{\gamma}\right)$ projects into the
$\pm$ band at momentum ${\bm k}$ and $\hat{\epsilon}_{\bm k} =
\vec{\epsilon}_{\bm k}/|\vec{\epsilon}_{\bm k}|$. The effect of
impurities on the normal state is accounted for by the renormalized
Matsubara frequencies 
\begin{equation}
\tilde{\omega}_{n,j} = \omega_n - \frac{1}{2\tau_{{\bm
    k},j}}\text{sgn}(\omega_n),
\end{equation}
 where the scattering rate (SR)
in band $j$ is given by
\begin{align}
\frac{1}{\tau_{{\bm k},j}} & = \sum_\alpha\pi n_{\text{imp},\alpha}
                             \left|V_\alpha\right|^2 \notag \\
& \phantom{=}\times\sum_{m=\pm}{\cal N}_{m}\left(1 +
jm\sum_{i=1}^5\phi_{\alpha}^{(i)}\hat{\epsilon}_{\bm{k},i}\langle \hat{\epsilon}_{\bm{k},i}\rangle_{m}\right),\label{eq:NormSR}
\end{align}
with ${\cal N}_{m}$ the density of states of band $m=\pm$ at the Fermi
surface, and
$\langle\ldots\rangle_{m}$ denotes the average over the Fermi surface
of this band. 
The second term in the parentheses of~\eq{eq:NormSR} arises from a net
average polarization in the internal degrees of freedom on the Fermi
surface of the $m$th band \cite{Cavanagh2020}. The factor
$\phi_{\alpha}^{(i)}=+ 1 (-1)$ if the scattering potential $\tilde{V}_\alpha$
commutes (anticommutes) with the term $\epsilon_{{\bm k},i}\gamma^i$
in the normal-state Hamiltonian~\eq{eq:genH}. This contribution can
have important effects on the overall scattering rate, enhancing or
reducing  the relative magnitude of interband to intraband
scattering. In the following we will assume a weak momentum-dependence
of the SR and replace $\tau^{-1}_{{\bf k},j}$ by its Fermi surface
average $\tau^{-1}_{j}$ in~\eq{eq:GF}.



\subsection{Superconducting properties}

In the orbital-spin basis, the pairing potential for a general superconducting state is $\Delta_{\bm k} = \Delta_0 \tilde{\Delta}_{{\bm k}}$ where  $\Delta_0$ is the magnitude and
\begin{equation}
 \tilde{\Delta}_{\bm k} = f_{\bm{k}}\gamma^{\alpha}\gamma^{\beta}U_T \,. \label{eq:genDelta}
\end{equation}
Here $f_{\bm{k}}$ is a normalized form factor, which must be chosen
such that fermionic antisymmetry is satisfied, i.e. $\tilde{\Delta}_{\bm
  k}=-\tilde{\Delta}^T_{-{\bm k}}$. The pairing potential
$\tilde{\Delta}_{\bm k}$ is a $4\times4$ matrix, and so there are only
six matrices defined by~\eq{eq:genDelta} for which an 
  even-parity form factor (i.e. $f_{\bm k}=f_{-\bm k}$) is permitted,
while the other ten matrices must have an odd-parity form factor (i.e. $f_{\bm
  k}=-f_{-\bm k}$). 

The pairing states which have $s$-wave form factor ($f_{\bm k}=1$) are
of central importance to our theory. 
 One such state is always given by $\alpha=\beta=0$ (where
 $\gamma^0=\mathbb{1}_4$), which describes the pairing of
   electrons in time-reversed-partner states, and is hence the 
 generalization 
 of the single-band conventional $s$-wave spin-singlet
 state. The remaining five $s$-wave channels depend
 nontrivially on the orbital degrees of freedom, i.e. 
 $\alpha\neq\beta$. The pairing potentials are determined by the form
 of the inversion operator: for a trivial inversion operator we have
 \begin{equation}
(\alpha,\beta) = (0,1),\, (0,2),\, (0,3),\, (0,4),\, (0,5)   \label{eq:trivialI}
 \end{equation}
whereas for a nontrivial inversion operator ${\cal I}=\gamma^1$ the
five potentials are
 \begin{equation}
(\alpha,\beta) = (0,1),\, (1,2),\, (1,3),\, (1,4),\, (1,5) \,. \label{eq:nontrivialI}
 \end{equation}
These additional $s$-wave channels may belong to nontrivial
irreps, and in general involve both intraband and interband
pairing. When projected onto the Fermi surface, these states will
typically have non-trivial momentum-dependence and gap nodes may be
present \cite{Cavanagh2020,Ramires2019}.

    
The degree to which the $s$-wave states involve interband pairing can be quantified by the
``superconducting fitness'' \cite{Ramires2018,Ramires2016}.
For convenience, we define the normalized superconducting fitness  on the $j$th band 
\begin{equation} 
\tilde{F}_C^{(j)} = \left\langle\frac{\text{Tr}\{|{\cal H}_{\bm k}\tilde{\Delta}_{\bm k}-\tilde{\Delta}_{\bm k}{\cal H}_{-\bm k}^T|^2\}}{|\vec{\epsilon}_{\bm  k}|^2\text{Tr}\{\tilde{\Delta}_{\bm k}\tilde{\Delta}^\dagger_{\bm k}\}}\right\rangle_{j}.\label{eq:bandFc}
\end{equation}
By definition, we have $0\leq \tilde{F}_C^{(j)}\leq 1$; this quantity is
related to the 
magnitude of the gap in the quasiparticle dispersion on the Fermi
surface $j$ by 
\begin{equation}
|\Delta_{{\bf k},j}| = \Delta_0\sqrt{1-\tilde{F}_C^{(j)}}.
\end{equation} 
When the pairing is purely intraband, the fitness
  $\tilde{F}_C^{(j)}$ is vanishing and the gap takes a maximal
value. Conversely, we see that gap nodes correspond to lines or points on
the Fermi surface where the gap is maximally unfit, i.e. the pairing
is purely interband. 

The fitness for the conventional $s$-wave state
is equal to zero, consistent with Anderson's theorem. The fitness for the unconventional
$s$-wave states in a system with trivial inversion symmetry
Eq.~\ref{eq:trivialI} evaluates as 
\begin{equation}
\tilde{F}_{C}^{(j)} = 1 - \langle \hat{\epsilon}^2_{{\bm k},\beta} \rangle_j\label{eq:triv_Fit}
\end{equation}
whereas in the system with nontrivial inversion
Eq.~\ref{eq:nontrivialI} the fitness for the unconventional $s$-wave
states is
\begin{equation}
\tilde{F}_{C}^{(j)} = \begin{cases}
1 - \langle \hat{\epsilon}^2_{{\bm k},1} \rangle_j & (\alpha,\beta)=(0,1)\\
\langle \hat{\epsilon}^2_{{\bm k},1} + \hat{\epsilon}^2_{{\bm k},n} \rangle_j &
(\alpha,\beta)=(1,n)
\end{cases}
\end{equation} 
Note that the odd-parity $s$-wave states typically have smaller
  values of $\tilde{F}_{C}^{(j)}$ than the even-parity $s$-wave states.

Expressed in a band-pseudospin basis, a pairing potential will
  typically have both inter- and intraband components. In particular,
  the intraband components will either be pseudospin-singlet or
  triplet, according as it is even or odd parity. Although the
  intraband pairing potential in the latter case
  is dependent upon the pseudospin basis, projecting the former into
  the pseudospin basis we find
\begin{equation}
\Delta_0\gamma^\beta U_T \rightarrow_{\pm} \begin{cases}
\Delta_0 is_y & \beta=0\\
\pm \Delta_0\hat{\epsilon}_{{\bf k},\beta}is_y & \beta\neq0
\end{cases}
\end{equation}
where $s_y$ is the Pauli matrix in the pseudospin degree of
freedom. In particular, we note that for the unconventional
($\beta\neq0$)  states the sign of the intraband potential
reverses between the two bands.

\subsection{The anomalous self-energy}

We now apply the Born approximation to the superconducting state,
specifically to determine the effect of the disorder on the critical
temperature. 
We distinguish between two cases: where pairing occurs in only a
single channel, and where multiple distinct pairing channels are
present. 

\subsubsection{Single channel} 

The pairing potential is determined self-consistently from the equation
\begin{equation}
\Delta_0 =
\frac{g_\nu}{2\beta}\sum_{i\omega_n}\int\!\!\frac{d^3k}{(2\pi)^3}\mbox{Tr}\{\tilde{\Delta}_{\bm
  k}^{\dagger} \bar{F}({\bf k},i\omega_n)\} \label{eq:gapeqn}
\end{equation}
where $g_{\nu}<0$ is the attractive interaction in a particular
superconducting channel $\nu$ and $\bar{F}({\bf k},i\omega_n)$ is the impurity-averaged anomalous
Green's function. In order to determine the critical temperature, we
expand the anomalous Green's function to linear order
in the pairing potential,
\begin{equation}
\bar{F}({\bf k},i\omega_n) \approx \bar{G}({\bm
  k},i\omega_n) ({\Delta}_{\bm k}+\Sigma_2)\bar{G}_{h}({\bm
    k},i\omega_n)
\end{equation}
 where $\bar{G}_h({\bm k},i\omega_n) = \bar{G}^{T}(-{\bm
   k},i\omega_n)$ is the impurity-averaged normal-state Green's
 function for the holes and $\Sigma_2$ is the anomalous self-energy
 due to the  impurity
scattering. Inserting this into Eq.~\ref{eq:gapeqn}, we obtain the linearized gap equation
\begin{align}
  \Delta_0 = & \frac{g_{\nu}}{2\beta}\sum_{i\omega_n}\int\!\!\frac{d^3k}{(2\pi)^3}\mbox{Tr}\{\tilde{\Delta}_{\bm k}^{\dagger}\bar{G}({\bm
  k},i\omega_n)\notag \\
& \times (\Delta_{\bm k}+\Sigma_2)\bar{G}_{h}({\bm
    k},i\omega_n)\} \label{eq:linearized}
\end{align}
The influence of disorder on the superconductivity is captured by the anomalous self-energy, which obeys the self-consistency condition
\begin{align}
\Sigma_2 = &
-\sum_{\alpha}n_{\text{imp},\alpha}\tilde{V}_\alpha\int\!\!\frac{d^3k}{(2\pi)^3}\bar{G}({\bm
  k},i\omega_n)\notag \\
& \times (\Delta_{\bm k}+\Sigma_2)\bar{G}_{h}({\bm k},i\omega_n)\tilde{V}^{T}_\alpha\,.\label{eq:Sigma2}
\end{align}
Importantly, the anomalous self-energy vanishes unless the lowest-order
contribution is nonzero: 
\begin{align}
  \Sigma^{(0)}_{2} &=  -\sum_{\alpha}n_{\text{imp},\alpha}\tilde{V}_\alpha\int\!\!\frac{d^3k}{(2\pi)^3}\bar{G}({\bm
  k},i\omega_n)\Delta_{\bm k}\bar{G}_{h}({\bm k},i\omega_n)\tilde{V}^{T}_\alpha
  \notag \\
  &=  \pi \sum_{\alpha}n_{\text{imp},\alpha}\tilde{V}_\alpha\sum\limits_{j=\pm}
  \frac{{\cal N}_j}{\left|\tilde{\omega}_{n,j}\right|} \langle {\cal P}_{{\bm
      k},j}\Delta_{\bm k}{\cal P}^T_{-{\bm
      k},j}\rangle_{j}\tilde{V}^{T}_\alpha. \label{eq:Sigma20}
\end{align}
In the final line of Eq. \ref{eq:Sigma20}, we have  neglected
  contributions to the self-energy due to interband pairing, on the
  the assumption that the  energy separation of the bands is much larger than the characteristic energy scales of the
superconductivity and the impurity scattering. 

For the two-band system considered here, the lowest-order contribution to the anomalous self-energy for  the general pairing state~\eq{eq:genDelta} is
\begin{widetext}
\begin{equation}
\Sigma^{(0)}_{2}  =  \sum_\alpha\pi n_{\text{imp},\alpha}\tilde{V}_\alpha\sum_{j=\pm}
\frac{{\cal N}_j}{4\left|\tilde{\omega}_{n,j}\right|} \Delta_0\left[\langle
                    f_{\bm k}\rangle_{j}\gamma^\alpha\gamma^\beta
                    + j\sum_{l=1}^{5} \langle f_{\bm
                    k}\hat{\epsilon}_{\bm{k},l}\rangle_{j} \left\lbrace
                    \gamma^\alpha\gamma^\beta,\gamma^l\right\rbrace  +
                    \sum_{l,m=1}^{5}\langle f_{\bm
                    k}\hat{\epsilon}_{\bm{k},l}\hat{\epsilon}_{\bm{k},m}\rangle_{j}\gamma^l\gamma^\alpha\gamma^\beta\gamma^m\right]U_T\tilde{V}^{T}_\alpha\,. \label{eq:s-wSig20}
\end{equation}
\end{widetext}
Since $\Sigma^{(0)}_2$ is an anomalous self-energy, is must satisfy the
fermionic antisymmetry condition 
\begin{equation}
\Sigma^{(0)}_2(i\omega_n) = -\Sigma^{(0)\,T}_2(-i\omega_n)\,.
\end{equation}
As seen from~\eqref{eq:s-wSig20}, the self-energy is even in frequency
and thus the only non-vanishing terms allowed in $\Sigma_2^{(0)}$ (and
therefore also $\Sigma_2$) are those proportional to the
unconventional $s$-wave potentials~\cite{Cavanagh2020}.

As was the case for the normal scattering rate, accounting for the
non-trivial structure of the disorder potentials necessitates the
inclusion of an additional parameter,
$\lambda_{\alpha}=+1(-1)$ if the gap 
 is fit (unfit) with respect to the scattering potential, i.e.   $\tilde{V}_\alpha\tilde{\Delta}_{\bm k}-\tilde{\Delta}_{\bm k}\tilde{V}_\alpha^T=0$ ($\neq 0$). For a
conventional $s$-wave gap, $\tilde{\Delta}=U_T$, this condition is exactly
equivalent to whether the disorder potential preserves or breaks
time-reversal symmetry. In analogy with the conventional case, for
unconventional $s$-wave states we will refer to scattering potentials
as `nonmagnetic' or `magnetic' according as they are fit or
  unfit with respect to the pairing potential. 

The lowest-order contribution to the anomalous self-energy for the
$s$-wave state $\nu$ can be expressed as
\begin{equation}
\Sigma_2^{(0)}=\sum_\alpha \pi n_{\text{imp},\alpha} \left|V_\alpha\right|^2\lambda_{\alpha} \sigma_{2,0} \tilde{\Delta}_\nu,
\end{equation}
where the form of $\sigma_{2,0}$ depends on the fitness functions 
\begin{equation}
\sigma_{2,0}=\frac{1}{2}\left[\mathcal{N}_{+}\frac{1-\tilde{F}_C^{(+)}}{\left|\omega_n\right|+\tau^{-1}_{+}}+\mathcal{N}_{-}\frac{1-\tilde{F}_C^{(-)}}{\left|\omega_n\right|+\tau^{-1}_{-}}\right].
\end{equation}
In particular, we observe that $\Sigma_2^{(0)}$ is only vanishing if
the fitness functions evaluate to $1$, i.e. the state corresponds to
purely interband pairing. Since such a situation is not
thermodynamically stable in the weak-coupling regime~\cite{Ramires2018}, it will be
generally true that an unconventional $s$-wave state with a nonzero
critical temperature has a nonzero anomalous self-energy.


\subsection{Multiple channels}


We consider a general pairing state
\begin{equation}
\Delta_{\bm{k}} =
\sum_{\mu}\Delta_0^{(\mu)}\tilde{\Delta}_{\mu,\bm
  k}
\end{equation}
where each channel $\mu$ belongs to the same irrep. The pairing
amplitudes $\Delta_0^{(\mu)}$ are fixed by solving the
self-consistency equations
\begin{equation}
\Delta_0^{(\mu)} = \sum_{\nu}\frac{g_{\mu,\nu}}{2\beta}\sum_{i\omega_n}\int\!\!\frac{d^3k}{(2\pi)^3}\mbox{Tr}\{\tilde{\Delta}_{\nu,\bm
  k}^{\dagger} \bar{F}({\bf k},i\omega_n)\}
\end{equation}
where $g_{\mu,\nu}$ is the pairing interaction which scatters a Cooper
pair in channel $\mu$ into channel $\nu$. To determine the critical
temperature we again linearize the anomalous Green's function to
obtain the linearized gap equation
\begin{align}
  \Delta_0^{(\mu)} = & \sum_{\nu}\frac{g_{\mu,\nu}}{2\beta}\sum_{i\omega_n}\int\!\!\frac{d^3k}{(2\pi)^3}\mbox{Tr}\{\tilde{\Delta}_{\nu,{\bm k}}^{\dagger}\bar{G}({\bm
  k},i\omega_n)\notag \\
& \times (\Delta_{\bm k}+\Sigma_2)\bar{G}_{h}({\bm
    k},i\omega_n)\}\,, \label{eq:linearized2}
\end{align}
where we have introduced the anomalous self-energy $\Sigma_2$ which is
determined self-consistently according to Eq.~(\ref{eq:Sigma2}). 

The anomalous self-energy will generally couple the various
superconducting channels to the $s$-wave states in the same irrep; we
see this explicitly in the lowest-order
contribution 
\begin{widetext}
\begin{align}
  \Sigma^{(0)}_{2} &=  \pi
                     \sum_{\mu}\sum_{\alpha}n_{\text{imp},\alpha}\tilde{V}_\alpha\sum\limits_{j=\pm}
                     \frac{{\cal
                     N}_j}{\left|\tilde{\omega}_{n,j}\right|} \langle
                     {\cal
                     P}_{\bm{k},j}\Delta_{0}^{(\mu)}\tilde{\Delta}_{\mu}{\cal
                     P}^T_{-\bm{k},j}\rangle_{j}\tilde{V}^{T}_\alpha=
                     \sum_{\mu}\sum_{\nu\in{s\text{-wave}}}
                     \Delta_{0}^{(\mu)} \sigma^{(\mu,\nu)}_{2,0} \tilde{\Delta}_\nu\label{eq:Sigma20_2comp}
\end{align}
\end{widetext}
where the indice $\mu$ ($\nu$) runs over the components of the order
parameter in all the (only the $s$-wave) channels. The contribution to the
self-energy in the $\nu$ $s$-wave channel due to the gap in the $\mu$ channel
is explicitly 
\begin{align}
\sigma_{2,0}^{(\mu,\nu)} = &\sum_\alpha
                     \pi n_{\text{imp},\alpha}
                     \left|V_\alpha\right|^2\lambda_{\alpha}^{(\nu)}\notag
                             \\
& \sum\limits_{j=\pm} \frac{{\cal
    N}_j}{4\left|\tilde{\omega}_{n,j}\right|}\text{Tr}\left\lbrace
  \left( \langle {\cal P}_{\bm{k},j}\tilde{\Delta}_{\mu,{\bf k}}{\cal P}^T_{-\bm{k},j}\rangle_{j}\right)\tilde{\Delta}_{\nu}^\dagger\right\rbrace.
\end{align}
where
$\lambda^{(\nu)}_{\alpha}=+1(-1)$  has the same meaning as in the single channel case, for each individual channel $\nu$.
The trace in this expression can be understood as a measure of the
overlap of the 
pairing state $\mu$ with the $s$-wave channel $\nu$ on the Fermi
surface $j$. 
The full anomalous self-energy is
\begin{equation}
\Sigma_2=\sum_{\mu,\nu}\Delta_{0}^{(\mu)}\sigma^{(\mu,\nu)}_{2} \tilde{\Delta}_\nu,
\end{equation}
with $\sigma_2$ given by the matrix equation
\begin{equation}
\sigma_2 = \left(\mathbb{1}-\sigma_{2,0}\right)^{-1}\sigma_{2,0},
\end{equation} 
where $\sigma_{2,0}$ has matrix elements $\sigma^{(\mu,\nu)}_{2,0}$,
which we take to be zero if $\nu$ does not correspond to an $s$-wave
channel. Inserting the self-energy into the linearized gap
equation~(\ref{eq:linearized2}), the critical temperature
is determined by the solution $\det\left\lbrace
  \mathcal{M}\right\rbrace =0$, with 
\begin{widetext}
\begin{align}
\mathcal{M}_{i,j} &= \delta_{i,j}-\sum_{\nu}\frac{g_{i,\nu}}{2\beta}\sum_{i\omega_n}\int\!\! \frac{d^3k}{(2\pi)^3}\mbox{Tr}\{\tilde{\Delta}_{\nu,\bm{ k}}^{\dagger}\bar{G}({\bm k},i\omega_n)[(\mathbb{1}-\sigma_{2,0})^{-1}]^{(i,j)}\tilde{\Delta}_{j,\bm{k}}\bar{G}_{h}({\bm k},i\omega_n)\}\,.
\end{align}
\end{widetext}
Note that in the following we will only consider the case where the
pairing interaction is diagonal, i.e. $g_{\nu,\mu}=g_{\nu}\delta_{\nu,\mu}$.

\section{Application to Dirac systems}\label{Sec:BiSe}

 We first consider the application of our formalism to a Dirac-like systems, as an example of the
   case where only a single
 band crosses the Fermi energy. For concreteness, we focus on
 the potential topological superconductor Cu$_x$Bi$_2$Se$_3$, making contact with previous work~\cite{MichaeliFu2012,Ramires2019,Sato2020,Cavanagh2020,Dentelski2020}. 

The low-energy electronic states in Bi$_2$Se$_3$ originate from the
outermost Se sites of the Bi$_2$Se$_3$ quintuple layers. These Se
sites are interchanged by inversion, and so give rise to a sublattice
structure. To
lowest order in $\bm{k}$ in each coefficient in Eq.~\ref{eq:genH}, the Hamiltonian  is given by~\cite{Liu2010_topins}
\begin{align}
  H =& -\mu \sigma_0\otimes\eta_0 + m \sigma_0\otimes\eta_x +
  v_zk_z\sigma_0\otimes\eta_y \notag \\
  & +v(k_x\sigma_y - k_y\sigma_x)\otimes\eta_z + \lambda
  k_x(k_x^2-3k_y^2)\sigma_z\otimes\eta_z \label{eq:CuBiSeHam}
\end{align}
where  $\sigma_\nu$ ($\eta_\nu$) are  Pauli matrices in spin (sublattice) space. The $\gamma$ matrices are defined
$\vec{\gamma} =
(\sigma_0\otimes\eta_x,\sigma_0\otimes\eta_y,\sigma_x\otimes\eta_z,\sigma_y\otimes\eta_z,\sigma_z\otimes\eta_z)$. 
The inversion symmetry operator is
$\mathcal{I}=\sigma_0\otimes\eta_x$; this term also appears
Hamiltonian as the mass term $m \sigma_0\otimes\eta_x$, which gaps out
the Dirac point at the Brillouin zone centre. The
copper intercalation only very weakly alters the bandstructure of the
topological insulator Bi$_2$Se$_3$, but dopes electrons into the
system, so that only  the upper band crosses the Fermi energy. 
For this system, there are four unconventional $s$-wave states
belonging to odd-parity irreps of the $D_{3h}$ point group, and an
additional unconventional $s$-wave state belonging to the trivial
$A_{1g}$ representation~\cite{FuBerg2010}. We tabulate the relevant signs of
$\lambda_{\alpha}$  for the $s$-wave states for all possible
impurity potentials in Table \ref{tab:CuxBi2Se3}.

\subsection{Odd-parity states}

 Solving the linearized gap equation~Eq.~\ref{eq:linearized}, the critical temperature of the odd-parity $s$-wave pairing states in
the presence of disorder is given by the solution of 
\begin{equation}
  \log\left(\frac{T_c}{T_{c0}}\right) = \psi\left(\frac{1}{2}\right) -
  \psi\left(\frac{1}{2} + \frac{1}{4\pi k_BT_c\tilde\tau_{\nu}}\right) 
\end{equation}
with the effective scattering rate 
is given by 
\begin{equation}
\tilde{\tau}_\nu^{-1} = \tau^{-1} - \sum_{\alpha}\pi\lambda_{\alpha} n_{\text{imp},\alpha} |V_\alpha|^2 \mathcal{N}\left(1-\tilde{F}_C\right), \label{eq:FAFC_v2_BiSe}
\end{equation}
and the normal-state scattering rate is
\begin{equation}
\tau^{-1}=\sum_\alpha \pi n_{\text{imp},\alpha} |V_\alpha|^2 \mathcal{N} (1+\phi_\alpha\langle
\hat{m}\rangle^2)
\end{equation}
Note that the mass term $m \sigma_0\otimes\eta_x$ generates a non-vanishing net
  spin-orbital polarization of the states at the Fermi
  surface~\cite{Cavanagh2020,MichaeliFu2012}. Since only the $+$ band
  crosses the Fermi energy, we drop the $+$ subscript on the density
  of states in these formulas. 

As can be readily seen in Table \ref{tab:CuxBi2Se3}, for each $s$-wave state there exists six
potentials (always including scalar disorder $\tilde V_\alpha = \mathbb{1}_4$) for which
$\lambda_{\alpha}=+1$ (``non-magnetic''), and the other ten potentials have
$\lambda_{\alpha}=-1$ (``magnetic'').  
If the disorder potential has $\lambda_{\alpha}=+1$, the effective SR is
reduced by an amount proportional to the degree of
fitness of the pairing potential: the fitter the gap (and therefore
the smaller $\tilde{F}_C$), the larger the reduction in the scattering
rate. 
In contrast, when $\lambda_{\alpha}=-1$ the SR is instead
enhanced, and the enhancement increases with increasing fitness.
Even if only ``nonmagnetic'' disorder is present, however, 
fine-tuning of the normal-state Hamiltonian is nevertheless required
for an odd-parity superconducting state to
be perfectly immune to disorder, since in general the superconducting
gap must be perfectly fit and the orbital-spin polarization of the
Fermi surface must be vanishing. 

An example of this fine-tuning has recently been provided in
  Ref. \cite{Dentelski2020}, which considers a purely Dirac system
with $m=0$, $v_z=v$ and $\lambda=0$ in Eq. \ref{eq:CuBiSeHam}. By
projecting the impurity potentials onto the band basis, it was
found that odd-parity superconducting states may be completely robust
against certain forms disorder. This result emerges
straightforwardly within our framework, where the complete
robustness is possible for the $A_{1u}$ gap since it
commutes with the remaining elements of the
Hamiltonian (proportional to the $\gamma^2$, $\gamma^3$, and
  $\gamma^4$ matrices)  and is
  therefore completely fit. For a
  general pairing potential, the effective scattering rate is 
\begin{equation}
\tilde{\tau}_\nu^{-1} = \sum_\alpha \pi n_{\text{imp},\alpha} |V_\alpha|^2 \mathcal{N}\left[1 - \lambda_{\alpha}\left(1-\tilde{F}_C\right)\right],
\label{eq:FAFC_v2_BiSe_Lim} 
\end{equation}
where the superconducting fitness is easily evaluated since the three
nonzero components of $\vec{\epsilon}_{\bf k}$ have equal magnitude on the (spherical) Fermi surface:
\begin{equation}
\tilde{F}_C=
\begin{cases} 0 & A_{1u} \\
1/3 & \text{other odd parity} 
\end{cases}
\end{equation}
From this the magnitude of the  effective scattering rates
given in table III of  Ref. \cite{Dentelski2020} follows immediately. 

\begin{table*}
\centering
  \begin{tabular}{|c|c||c|c|c|c|c|c|c|c|c|c|c|c|c|c|c|c|}\hline
   & & \multicolumn{16}{c|}{$\tilde{V}_{\alpha}$}\\\hline
   & &  $\gamma^0$& $\gamma^1$& $\gamma^2$& $\gamma^3$& $\gamma^4$& $\gamma^5$ & $i\gamma^1\gamma^2$ & $i\gamma^1\gamma^3$ & $i\gamma^1\gamma^4$ & $i\gamma^1\gamma^5$ & $i\gamma^2\gamma^3$ & $i\gamma^2\gamma^4$ & $i\gamma^2\gamma^5$ & $i\gamma^3\gamma^4$ & $i\gamma^3\gamma^5$& $i\gamma^4\gamma^5$ \\\hline
  irrep & $\tilde{\Delta}U_T^\dagger$ & $\mathbb{I}_4$ & $\sigma_0\eta_x$& $\sigma_0\eta_y$& $\sigma_x\eta_z$& $\sigma_y\eta_z$& $\sigma_z\eta_z$& $\sigma_0\eta_z$ & $\sigma_x\eta_y$ & $\sigma_y\eta_y$ & $\sigma_z\eta_y$& $\sigma_x\eta_x$ & $\sigma_y\eta_x$ & $\sigma_z\eta_x$& $\sigma_z\eta_0$ & $\sigma_y\eta_0$ & $\sigma_x\eta_0$ \\\hline
$A_{1g}$ & $\gamma^0$ &
\lamp & \lamp & \lamn & \lamn & \lamn & \lamn & \lamp & \lamp & \lamp & \lamp & \lamn & \lamn & \lamn & \lamn & \lamn & \lamn \\\hline
$A_{1g}$ & $\gamma^1$ & \lamp & \lamp & \lamp & \lamp & \lamp & \lamp & \lamn & \lamn & \lamn & \lamn & \lamn & \lamn & \lamn & \lamn & \lamn & \lamn \\\hline
$A_{1u}$ & $i\gamma^1\gamma^5$ & \lamp & \lamn & \lamn & \lamn & \lamn & \lamp & \lamn & \lamn & \lamn & \lamp & \lamn & \lamn & \lamp & \lamn & \lamp & \lamp \\\hline
$A_{2u}$ & $i\gamma^1\gamma^2$ & \lamp & \lamn & \lamp & \lamn & \lamn & \lamn & \lamp & \lamn & \lamn & \lamn & \lamp & \lamp & \lamp & \lamn & \lamn & \lamn \\\hline
\multirow{2}{*}{$E_{u}$} & $i\gamma^1\gamma^3$ &
                                                 \lamp & \lamn
& \lamn & \lamp & \lamn & \lamn & \lamn & \lamp & \lamn & \lamn & \lamp & \lamn & \lamn & \lamp & \lamp & \lamn\\ \hhline{|~|-----------------|}
 & $i\gamma^1\gamma^4$ & \lamp & \lamn & \lamn & \lamn & \lamp & \lamn & \lamn & \lamn & \lamp & \lamn & \lamn & \lamp & \lamn & \lamp & \lamn & \lamp \\\hline
  \end{tabular}
  \caption{The value of $\lambda_{\alpha}$ for the
    sixteen  possible momentum-independent disorder potentials, for
    each of the six possible unconventional $s$-wave states, for
    Cu$_x$Bi$_2$Se$_3$.  This overall sign is calculated from
    $\tilde{V}_\alpha\tilde{\Delta}U_T\tilde{V}^{\text{T}}_\alpha =
    \lambda_{\alpha}\tilde{\Delta}U_T$. Disorder potentials
    with $\lambda_{\alpha}=-1$ for the conventional
    $A_{1g}$ gap ($\gamma^0$) break time-reversal symmetry. The
      ``$\otimes$'' in the orbital-spin form of the pairing and
      impurity potentials is omitted for clarity. \label{tab:CuxBi2Se3}}
\end{table*}

\subsection{Even-parity states}

 To conclude this section, we note that the analysis for the
  $s$-wave $A_{1g}$ states is somewhat more complicated since the
  anomalous self-energy will generally always contain terms
  proportional to the conventional and unconventional pairing
  potentials. The general form for the $A_{1g}$ gap is
  \begin{equation}
\Delta = \left[\Delta_{0}^{(0)}\gamma^{0}+ \Delta_{0}^{(1)}\gamma^{1}\right]U_T,\label{eq:BiSe_2comp_gap}
\end{equation}
and the corresponding lowest-order contribution to the anomalous self-energy  is given by
\begin{widetext}
\begin{align}
\Sigma^{(0)}_{2}  &= \sum_\alpha\pi n_{\text{imp},\alpha}|\tilde{V}_\alpha|^2 \frac{{\cal N}}{2\left|\tilde{\omega}_{n}\right|} \begin{pmatrix}
\Delta_0^{(0)} \\
\Delta_0^{(1)}
\end{pmatrix}^{\text{T}}
\begin{bmatrix}
\lambda^{(0)}_\alpha &   \lambda^{(1)}_\alpha\langle \hat{m}\rangle \\
\lambda^{(0)}_\alpha \langle\hat{m}\rangle & \lambda^{(1)}_\alpha\langle\hat{m}^2\rangle
\end{bmatrix}
\begin{pmatrix}
\gamma^0 \\
\gamma^1
\end{pmatrix}U_T\label{eq:Sig20_BiSe_A1g}
\end{align}
\end{widetext}
Because there is only a single band at the Fermi level, the effective scattering rate has the form
\begin{equation}
\tilde{\tau}_\nu^{-1} = \tau^{-1} - \sum_{\alpha}\pi
n_{\text{imp},\alpha}|V_\alpha|^2 {\cal N}\left(\lambda^{(0)}_\alpha +
  \lambda_{\alpha}^{(1)} \langle \hat{m}\rangle^2 \right)
\end{equation}
where $\lambda_\alpha^{(0)}$ and $\lambda_{\alpha}^{(1)}$ are the
$\lambda$-factors for the conventional and unconventional states,
respectively. We observe that 
$\phi_\alpha=\lambda_\alpha^{(0)}\lambda_{\alpha}^{(1)}$, from which
it follows that the effective scattering rate is vanishing for
$\lambda_{\alpha}^{(0)}=1$, i.e. the $s$-wave $A_{1g}$ state is
insensitive to disorder which preserves time-reversal symmetry,  as
required by Anderson's theorem. The complete robustness of the
unconventional $s$-wave state is due to the fact that the two gaps are
indistinguishable on the single Fermi
surface. As we will demonstrate
in the following section, when both bands cross the Fermi energy the general gap 
becomes sensitive to
time-reversal symmetry preserving disorder due to the unconventional
component. This does not, however, violate Anderson's theorem as
the unconventional component does not pair time-reversed partners.

\section{Application to the iron pnictides}\label{Sec:FeAs}

The framework we have presented can be applied in a straightforward way to systems with considerably more complicated Fermi surfaces than the single sheet Fermi surfaces of Dirac-like materials. To highlight this generality, we apply our method to a model for the iron pnictide superconductors for which two bands cross the Fermi level, each contributing two sheets to the Fermi surface.

We use a tight-binding model of the iron oxypnictides proposed by Raghu
\emph{et al}. \cite{Raghu2008}, which includes only the
  contribution from the iron $d_{xz}$ and $d_{yz}$ orbitals. 
More sophisticated models, including up to 
five or more orbitals \cite{Kuroki2008,
  Eschrig2009, Graser2010} 
better reproduce the
electronic structure, but our focus here is in
understanding the influence of the multiple Fermi surfaces,
for which Raghu's model is sufficient.  
The Hamiltonian for Raghu's model is written
\begin{align}
\mathcal{H}_{\bm{k}} &= \varepsilon_0\left(\bm{k}\right) \sigma_0\otimes\tau_0 +\varepsilon_z\left(\bm{k}\right) \sigma_0\otimes\tau_z +\varepsilon_x\left(\bm{k}\right) \sigma_0\otimes\tau_x\nonumber\\
&\qquad\qquad+\lambda \sigma_z\otimes\tau_y \nonumber\\
&= \varepsilon_0\left(\bm{k}\right) \gamma^0 +\varepsilon_z\left(\bm{k}\right)\gamma^1+\varepsilon_x\left(\bm{k}\right)\gamma^2+\lambda\gamma^5,
\end{align}
with 
$\varepsilon_0 = -\mu
-(t_1+t_2)[\cos(k_x)+\cos(k_y)]-4t_3\cos(k_x)\cos(k_y) $,
$\varepsilon_x=-2t_4\sin(k_x)\sin(k_y)$ and
$\varepsilon_z=-(t_1-t_2)[\cos(k_x)-\cos(k_y)]$. The Pauli
matrices $\tau_\nu$ encode the iron $d_{xz}$ and $d_{yz}$ orbital
degree of freedom, which transform trivially under inversion
($\mathcal{I}=\mathbb{1}$) and time-reversal. We extend the original
model of Ref. \cite{Raghu2008} by including an additional spin-orbit
coupling, in keeping with more general proposals
\cite{Cvetkovic2013,Vafek2017}. Throughout, we use
the parameters $\left\lbrace t_1, t_2, t_3, t_4,
  \mu\right\rbrace=\left\lbrace -1, 1.3, -0.85, -0.85, 1.45
\right\rbrace\left|t_1\right|$~ \cite{Raghu2008},
and examine a variety of magnitudes of the spin-orbit coupling $\lambda$.

The normal-state scattering rates on the two bands are in general
different and given by
\begin{align}
\tau_{\pm}^{-1} = & \sum_\alpha\frac{\pi
                  n_{\text{imp},\alpha}\left|V_\alpha\right|^2}{2}\left[\mathcal{N}_{\pm}\left(1+\phi_{\alpha}\left\langle\hat{\lambda}\right\rangle_{\pm}^2\right)\right.
                  \notag \\
& \left. +\mathcal{N}_{\mp}\left(1-\phi_{\alpha} \left\langle\hat{\lambda}\right\rangle_{\pm}\left\langle\hat{\lambda}\right\rangle_{\mp}\right)\right]\,.
\end{align}
The second term in the brackets is the
 contribution from interband scattering. The
  non-zero Fermi surface average of the spin-orbit coupling gives a
  non-trivial dependence on the impurity potential. Specifically, 
for $\phi_{\alpha}=+1$, the spin-orbit coupling 
enhances intraband scattering and suppresses interband scattering,
whereas $\phi_{\alpha}=-1$ gives the opposite effect.
 This result can be easily understood in the extreme limit
$\hat{\lambda}\rightarrow 1$ where the two bands become
eigenstates of the spin-orbit coupling operator $\gamma^5$. 
  Scattering off a disorder potential which commutes with $\gamma^5$
(and therefore has $\phi_{\alpha}=1$) does not change
  the $\gamma^5$-eigenstate of the electron, and hence cannot scatter
  between the two bands. Similarly, disorder that anti-commutes with
  the spin-orbit coupling operator is incapable of intraband
  scattering in this extreme limit. On the other hand, when
  $\lambda=0$ there is no net spin-orbital
  polarization of either band, and the scattering rates in the two
  bands become indistinguishable. 

Due to the trivial inversion symmetry, the unconventional
$s$-wave states have even parity. As tabulated in
  Tab.~\ref{tab:Raghu_imp}, the unconventional $s$-wave states belong to
  the $A_{1g}$, $B_{1g}$, $B_{2g}$ and $E_g$
irreps of the $D_{4h}$ point group. From Eq.~\ref{eq:triv_Fit} we observe that the
$E_g$ gaps are completely unfit since the two-dimensional Hamiltonian does not contain
any terms proportional to $\gamma^3$ and $\gamma^4$, and so we will
not consider the $E_g$ states in the following.
Since the $s$-wave pairing states are all even-parity, their
projections into a pseudospin basis is explicitly given by
\begin{eqnarray}
\Delta_0\gamma^{0}U_T & \rightarrow & \Delta_\pm = \Delta_0is_y\\
\Delta_0\gamma^{1}U_T & \rightarrow & \Delta_\pm = \pm \hat{\epsilon}_z\Delta_0is_y\\
\Delta_0\gamma^{2}U_T & \rightarrow & \Delta_\pm = \pm \hat{\epsilon}_x\Delta_0is_y\\
\Delta_0\gamma^{5}U_T & \rightarrow& \Delta_\pm = \pm \hat{\lambda}\Delta_0is_y
\end{eqnarray}
Note that the unconventional $A_{1g}$ state has an $s^{\pm}$ form,
with a full gap with opposite signs on the electron- and hole-like
Fermi surfaces. This is widely accepted as the sign structure of the
pairing state in the iron pnictides, although it is important to
emphasize that this can be achieved with an orbitally-trivial pairing
potential, as we discuss below.

\ignore{
\begin{table}
\centering
   \begin{tabular}{|c||c|c|c|c|c|c|}\hline
     irrep & $A_{1g}$ & $B_{1g}$ & $B_{2g}$ & \multicolumn{2}{c|}{$E_{g}$}& $A_{1g}$ \\\hline
    $\tilde{\Delta}U_T^\dagger$ & $\mathbb{I}_4$ & $\gamma^1$& $\gamma^2$&
    $\gamma^3$& $\gamma^4$ & $\gamma^5$\\\hline
   nodes & none & line & line & line & line & none\\\hline
     $l$, $\lambda_{l}=1$ & all & 1 & 2 & 3 & 4 & 5\\\hline
     $l$, $\lambda_{l}=-1$ & none & 2,3,4,5 & 1,3,4,5 & 1,2,4,5 & 1,2,3,5 & 1,2,3,4 \\\hline
  \end{tabular}
  \caption{The six $s$-wave pairing states for Raghu's model of the iron oxypnictides. The first line
    gives the irrep of $D_{4h}$, the second line gives the form of the pairing
    potential in terms of the $\gamma$ matrices defined in the text,
    the third line gives the nodal structure, 
    and the final two lines give the
    values of $l$ corresponding to the $\gamma$ matrices for which 
    $\lambda_{l}=1$ and
    $\lambda_{l}=-1$, respectively. Terms in the normal state Hamiltonian with $\lambda_l=+1$ reduce the superconducting fitness (i.e. the gap is fit with respect to these terms), while those with  $\lambda_l=-1$ increase the superconducting fitness, and therefore reduce the robustness for disorder with $\lambda_{\text{imp}}=+1$. \label{tab:Raghu}}
\end{table}
}

\begin{table*}
\centering
  \begin{tabular}{|c|c||c|c|c|c|c|c|c|c|c|c|c|c|c|c|c|c|}\hline
   & & \multicolumn{16}{c|}{$\tilde{V}_{\alpha}$}\\\hline
   & &  $\gamma^0$& $\gamma^1$& $\gamma^2$& $\gamma^3$& $\gamma^4$& $\gamma^5$ & $i\gamma^1\gamma^2$ & $i\gamma^1\gamma^3$ & $i\gamma^1\gamma^4$ & $i\gamma^1\gamma^5$ & $i\gamma^2\gamma^3$ & $i\gamma^2\gamma^4$ & $i\gamma^2\gamma^5$ & $i\gamma^3\gamma^4$ & $i\gamma^3\gamma^5$& $i\gamma^4\gamma^5$ \\\hline
  irrep & $\tilde{\Delta}U_T^\dagger$ & $\mathbb{I}_4$ & $\sigma_0\eta_z$& $\sigma_0\eta_x$& $\sigma_x\eta_y$& $\sigma_y\eta_y$& $\sigma_z\eta_y$& $\sigma_0\eta_y$ & $\sigma_x\eta_x$ & $\sigma_y\eta_x$ & $\sigma_z\eta_x$& $\sigma_x\eta_z$ & $\sigma_y\eta_z$ & $\sigma_z\eta_z$& $\sigma_z\eta_0$ & $\sigma_y\eta_0$ & $\sigma_x\eta_0$ \\\hline
$A_{1g}$ & $\gamma^0$ & \lamp & \lamp & \lamp & \lamp & \lamp & \lamp & \lamn & \lamn & \lamn & \lamn & \lamn & \lamn & \lamn & \lamn & \lamn & \lamn \\\hline
$B_{1g}$ & $\gamma^1$ & \lamp & \lamp & \lamn & \lamn & \lamn & \lamn & \lamp & \lamp & \lamp & \lamp & \lamn & \lamn & \lamn & \lamn & \lamn & \lamn \\\hline
$B_{2g}$ & $\gamma^2$ & \lamp & \lamn & \lamp & \lamn & \lamn & \lamn & \lamp & \lamn & \lamn & \lamn & \lamp & \lamp & \lamp & \lamn & \lamn & \lamn \\\hline
\multirow{2}{*}{$E_{g}$} & $\gamma^3$ & \lamp & \lamn & \lamn & \lamp & \lamn & \lamn & \lamn & \lamp & \lamn & \lamn & \lamp & \lamn & \lamn & \lamp & \lamp & \lamn\\ \hhline{|~|-----------------|}
 & $\gamma^4$ & \lamp & \lamn & \lamn & \lamn & \lamp & \lamn & \lamn & \lamn & \lamp & \lamn & \lamn & \lamp & \lamn & \lamp & \lamn & \lamp \\\hline
$A_{1g}$ & $\gamma^5$ & \lamp & \lamn & \lamn & \lamn & \lamn & \lamp & \lamn & \lamn & \lamn & \lamp & \lamn & \lamn & \lamp & \lamn & \lamp & \lamp \\\hline
  \end{tabular}
  \caption{The value of $\lambda_{\alpha}$ for the sixteen  possible momentum-independent disorder potentials, for each of the six possible unconventional $s$-wave states for the model of iron pnictide superconductors proposed by Raghu \emph{et al.} \cite{Raghu2008}.  Disorder potentials with $\lambda_{\alpha}=-1$ for the first $A_{1g}$ gap, proportional to the identity matrix, break time-reversal symmetry. \label{tab:Raghu_imp}}
\end{table*}

\subsection{The $B_{1g}$ and $B_{2g}$ irreps}

We begin by considering the $B_{1g}$ and $B_{2g}$ irreps, which both
have a single unconventional $s$-wave pairing potential. We first examine the
robustness of the unconventional $s$-wave state, and then compare this
against orbitally-trivial $d$-wave spin-singlet pairing states in the
same irrep. 

\subsubsection{$s$-wave gaps}


The usual Abrikosov-Gor'kov result for the suppression of an
unconventional pairing state in a two-band system is
\begin{equation}
  \log\left(\frac{T_c}{T_{c0}}\right) = \sum_{j= \pm}R_j\left[\psi\left(\frac{1}{2}\right) -   \psi\left(\frac{1}{2} + \frac{1}{4\pi k_BT_c\tau_{j}}\right)\right] \label{eq:AG2band},
\end{equation}
where the contribution of the $j$th band is weighted according to its
contribution to the condensation energy
\begin{equation}
R_j = \frac{{\cal N}_j\langle |\tilde \Delta_{\bm k}|^2\rangle_{j} }{{\cal
               N}_+ \langle |\tilde \Delta_{\bm k}|^2\rangle_{+} + {\cal
               N}_- \langle |\tilde \Delta_{\bm k}|^2\rangle_{-}}\,. \label{eq:R0}
\end{equation}
This result naturally reduces to the single-band case in the limit
where one of the densities of states vanishes. Moreover, in the
absence of the spin-orbit coupling the normal-state scattering rates
are independent of the band index, i.e. 
$\tau_{\pm}^{-1}=\tau^{-1}$, and we recover the usual universal
result. 

\begin{figure}
	\begin{overpic}[width=\columnwidth]{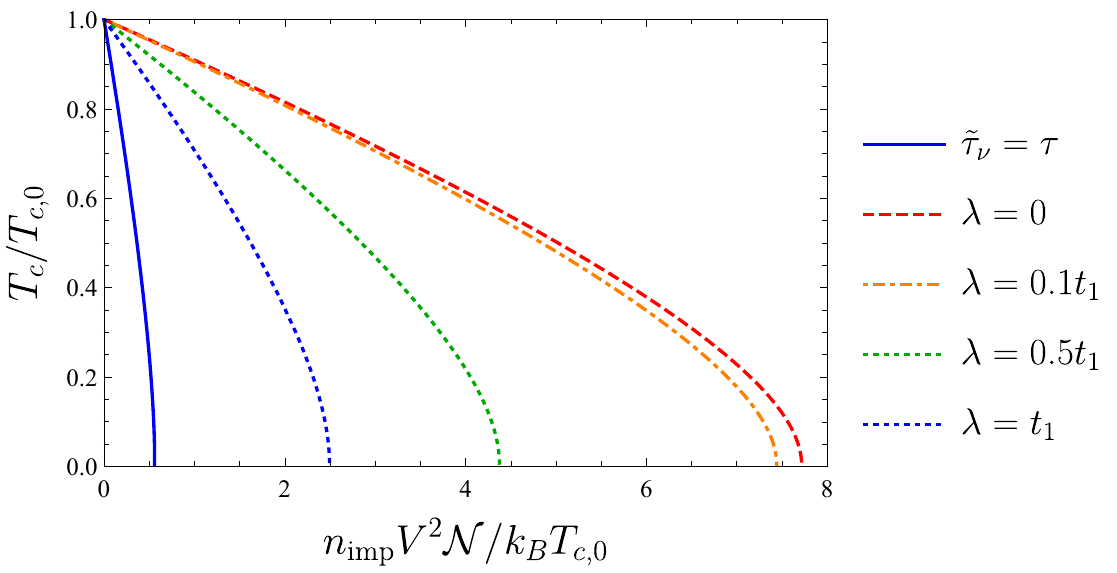}
	\put (50,45) {$B_{1g}$}
	\end{overpic}\\
	\begin{overpic}[width=\columnwidth]{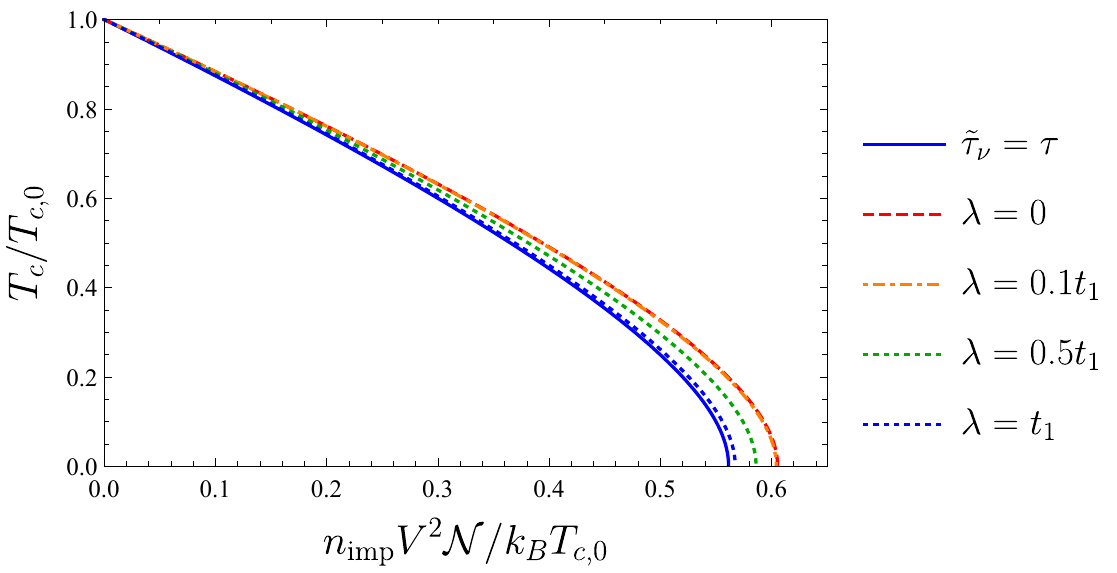}
	\put (50,45) {$B_{2g}$}
	\end{overpic}
  \caption{Robustness against ``nonmagnetic'' ($\lambda_{\alpha}=+1$)
    disorder for the $B_{1g}$ (top) and $B_{2g}$ (bottom) $s$-wave
    gaps in the two-orbital model of the iron pnictides, for 
      various values of the spin-orbit coupling $\lambda$.\label{fig:FeAs_BgCurves}}
\end{figure}

Accounting for the nontrivial orbital-spin structure of the
unconventional $s$-wave pairing state in channel $\nu$, the expression Eq.~\ref{eq:AG2band} is modified as
$\tau_{j}^{-1} \rightarrow \bar{\tau}_{\nu,j}^{-1}$ and
$R_j\rightarrow \bar{R}_j$. The effective scattering rate
$\bar{\tau}_{\nu,j}^{-1}$ is  given
by  
\begin{widetext}
\begin{align}
\bar{\tau}^{-1}_{\nu,\pm} &=
                            \frac{\tilde{\tau}^{-1}_{\nu,+}+\tilde{\tau}^{-1}_{\nu,-}}{2}\pm \sqrt{\left[\frac{\tilde{\tau}^{-1}_{\nu,+}-\tilde{\tau}^{-1}_{\nu,-}}{2}\right]^2 + \left(\tilde{\tau}^{-1}_{\nu,+}-\tau^{-1}_{+}\right)\left(\tilde{\tau}^{-1}_{\nu,-}-\tau^{-1}_{-}\right)}, \label{eq:SReff_2FS}
\end{align}
\end{widetext}
with
\begin{equation}
\tilde{\tau}^{-1}_{\nu,j} = \tau^{-1}_j - \sum_{\alpha}\pi\lambda_{\alpha} n_{\text{imp},\alpha} |V_\alpha|^2 \mathcal{N}_j\left(1-\tilde{F}^{(j)}_C\right)\,. \end{equation}
We recognize the $\tilde{\tau}_{\nu,j}^{-1}$ as the generalization of
Eq.~\ref{eq:FAFC_v2_BiSe} to the multiband case, assuming that we can treat each band
independently. However, the unconventional $s$-wave pairing potentials
couple the two bands, and we therefore cannot readily associate the
scattering rates in 
Eq.~\ref{eq:SReff_2FS}  with one band or
the other. 
\ignore{ if we treated each band independently. However, the effective
  SRs $\bar{\tau}^{-1}_{\nu,\pm}$ which
  appear in~Eq. \ref{eq:modifiedAG} cannot be unambiguously associated with a
  single band, reflecting the presence of
  interband scattering.}
The weighting of the contribution from the two effective scattering
rates also deviates from the expected form Eq.~\ref{eq:R0}, and depends
upon the normal-state and effective scattering rates
\begin{widetext}
\begin{align}
\bar{R}_{\pm}&= R_{\pm} \ignore{\frac{{\cal
               N}_\pm \langle |\tilde \Delta_{\bm k}|^2\rangle_{\pm}}{{\cal
               N}_+ \langle |\tilde \Delta_{\bm k}|^2\rangle_{+} + {\cal
               N}_- \langle |\tilde \Delta_{\bm k}|^2\rangle_{-}}} \pm  \frac{{\cal
               N}_+ \langle |\tilde \Delta_{\bm k}|^2\rangle_{+}(\bar{\tau}_{\nu,-}^{-1} -
               \tau_{-}^{-1}) + {\cal
               N}_- \langle |\tilde \Delta_{\bm k}|^2\rangle_{-}(\bar{\tau}_{\nu,+}^{-1} -
               \tau_{+}^{-1})}{[{\cal
               N}_+ \langle |\tilde \Delta_{\bm k}|^2\rangle_{+} + {\cal
               N}_- \langle |\tilde \Delta_{\bm
               k}|^2\rangle_{-}](\bar{\tau}_{\nu,+}^{-1}-\bar{\tau}_{\nu,-}^{-1})} \label{eq:barR}
\end{align}
\end{widetext}
Using these expressions, we plot the critical temperature of the $B_{1g}$ and
$B_{2g}$ $s$-wave gaps as a function of the strength of 
``nonmagnetic'' and ``magnetic'' disorder  in
Figs. \ref{fig:FeAs_BgCurves} and \ref{fig:FeAs_Bg_TRSB},
respectively. Notably, the $B_{1g}$ state is extremely robust
against ``nonmagnetic'' disorder, with superconductivity persisting to a
disorder strength more than ten times that expected
from~Eq.~\ref{eq:AG2band}; in contrast, the $B_{2g}$ state closely 
follows the predictions of the Abrikosov-Gor'kov theory.

\ignore{
The robustness of the $B_{1g}$ and $B_{2g}$ $s$-wave gaps against
``nonmagnetic'' and ``magnetic'' disorder are shown in
Figs. \ref{fig:FeAs_BgCurves} and \ref{fig:FeAs_Bg_TRSB},
respectively. Notably, the $B_{1g}$ state is \emph{extremely} robust
against ``nonmagnetic'' disorder, with superconductivity persisting to a
disorder strength more than ten times that expected for an
unconventional pairing state in a single-band model with comparable; in contrast, the $B_{2g}$ state closely
resembles the result for a 
}

The pronounced robustness of the $B_{1g}$ state is due to the
almost-perfect fitness on the Fermi surface of the $+$ band. For
realistic values of the spin-orbit coupling, on this Fermi
  surface we have 
$\varepsilon_{z}\gg\varepsilon_x$, $\lambda$, and so  $F_{C}^{(+)}
= 1 - \langle \hat{\varepsilon}_z^2\rangle_{+} \approx 0$; on the other hand, the average fitness on
the Fermi surfaces of the $-$ band is much smaller, with the gap
displaying 
nodes along the Brillouin zone diagonals where the potential is
completely unfit. In the limit where the spin-orbit coupling is
vanishing, we find that the effective scattering rates are
\begin{equation}
\bar{\tau}_{\nu,\pm}^{-1} = \frac{\pi}{4}\sum_j\sum_\alpha
n_{\text{imp},\alpha}|V_\alpha|^2{\cal N}_j(2 \pm [1 \mp \lambda_{\alpha}][1-F_C^{(j)}])\,.
\end{equation}
Taking
$F_{C}^{(+)}\approx 0$, and restricting to ``nonmagnetic'' disorder
($\lambda_{\alpha}=+1$) we find that
$\bar{\tau}^{-1}_{\nu,+}\approx \tau^{-1}$ and
$\bar{\tau}^{-1}_{\nu,-}\approx \pi\sum_\alpha
n_{\text{imp},\alpha}|V_\alpha|^2{\cal N_-}F_C^{(-)}$. Since
${\cal N}_+ \gg {\cal N}_-$ in Raghu's model, we expect that
$\bar{\tau}^{-1}_{\nu,-} \ll \tau^{-1}$. Inserting these
expressions into Eq.~\ref{eq:barR} we find that $\bar{R}_+\approx 0$
and $\bar{R}_-\approx 1$. We thus see that critical temperature is
controlled by
one effective SR $\bar{\tau}^{-1}_{\nu,-}$, which can be much smaller
than the normal-state SR. Note that due to the almost-perfect fitness
of the superconductivity on the Fermi surfaces of the $+$ band, the
disorder-response is completely controlled by the fitness on the Fermi
surfaces of the $-$ band.

The comparable fragility of the $B_{2g}$ gap can also be understood
using these arguments. The $B_{2g}$ gap is much less fit than the
$B_{1g}$ gap on the Fermi surface of the $+$ band, but has comparable
fitness on the Fermi surfaces of the $-$ band. Repeating the analysis
above, but now taking $\tilde{F}_C^{(+)}\approx 1$, we find that $\bar{\tau}^{-1}_{\nu,+}\approx \tau^{-1}$ and
$\bar{\tau}^{-1}_{\nu,-}\approx \tau^{-1} - \pi\sum_\alpha
n_{\text{imp},\alpha}|V_\alpha|^2{\cal N_-}[1-F_C^{(-)}]$. For
the parameters of Raghu's model, the two effective scattering rates
are similar, and the suppression of the pairing by disorder is 
therefore well 
approximated by the Abrikosov-Gor'kov result. 



\begin{figure}
\centering
	\begin{overpic}[width=0.5\columnwidth]{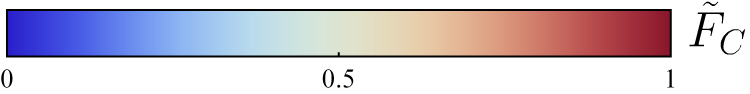}
	\end{overpic}\\
	\begin{overpic}[width=0.45\columnwidth]{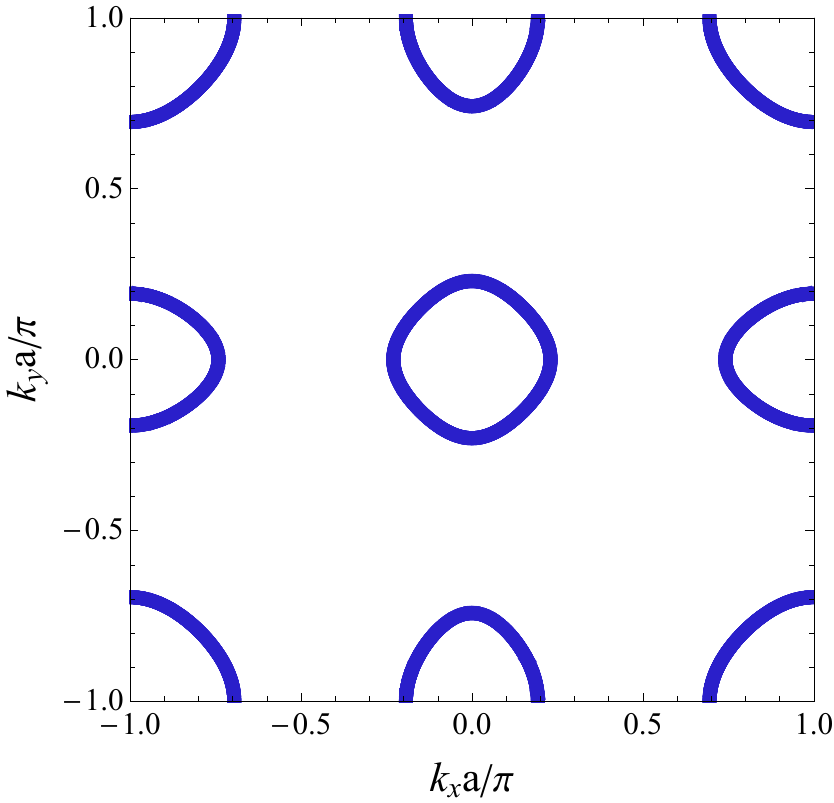}
	\put (20,65) {$A_{1g}$, $\gamma^0$}
	\end{overpic}
	\begin{overpic}[width=0.45\columnwidth]{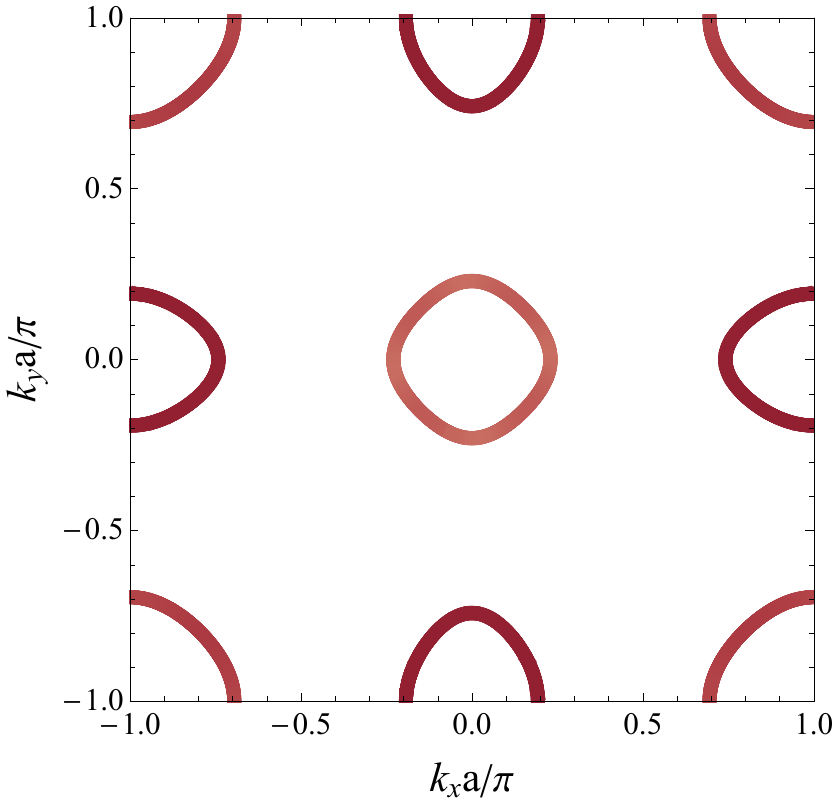}
	\put (20,65) {$A_{1g}$, $\gamma^5$}
	\end{overpic}\\
	\begin{overpic}[width=0.45\columnwidth]{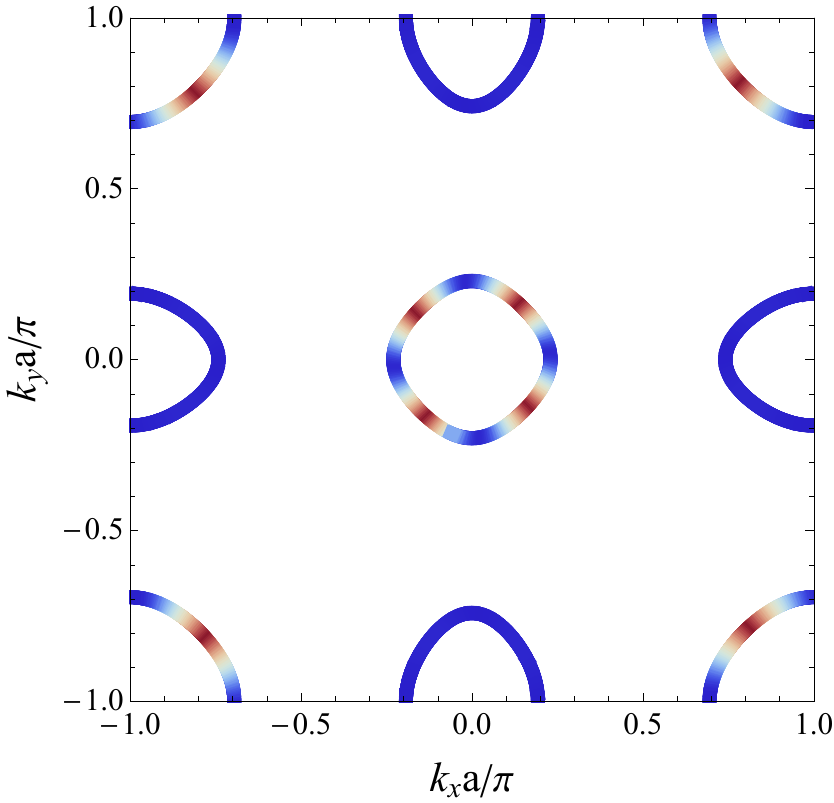}
	\put (20,65) {$B_{1g}$}
	\end{overpic}
	\begin{overpic}[width=0.45\columnwidth]{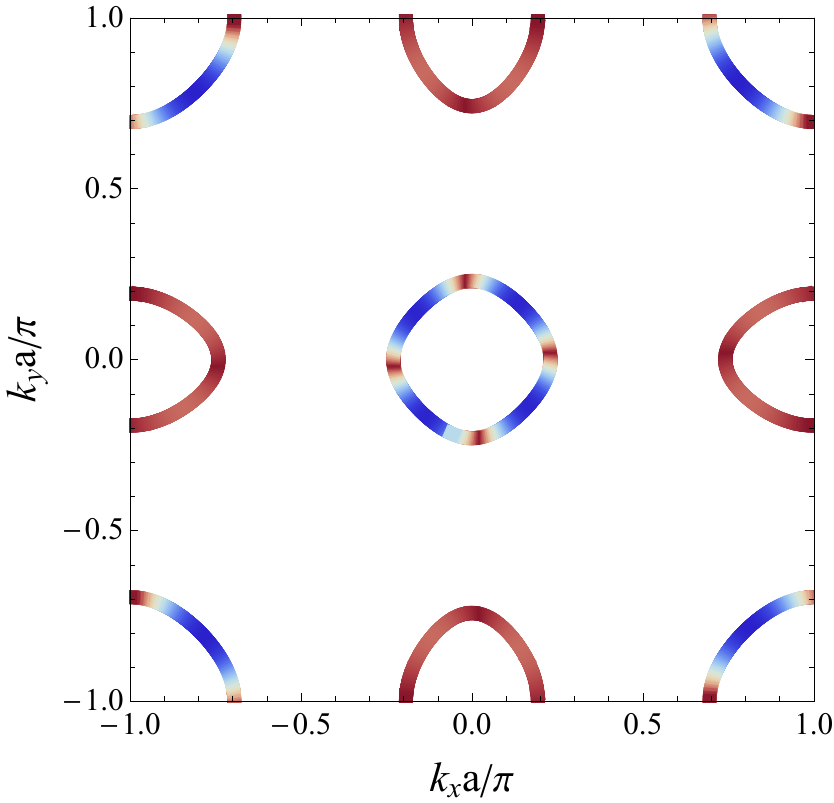}
	\put (20,65) {$B_{2g}$}
	\end{overpic}
  \caption{Fermi surface for the two-orbital model
    for iron pnictides, with $\lambda=0.1\left|t_1\right|$, and superconducting fitness for the four $s$-wave states considered. The coloring of the contours denotes the magnitude of the superconducting fitness. For the conventional $A_{1g}$ state ($\gamma^0$), the fitness is zero everywhere, while for the unconventional state ($\gamma^5$) it varies between the bands. The $B_{1g}$ state also has vanishing averaged fitness on the $+$ band Fermi surface, while the $B_{2g}$ fitness is finite everywhere on the Fermi surface. \label{fig:FeAs_FS_OrbitalContent} }
\end{figure}

\ignore{The robustness of the $B_{1g}$ $s$-wave gap can be understood on the
basis of the orbital content of the two bands, as shown in
Fig. \ref{fig:FeAs_FS_OrbitalContent}. The $B_{1g}$ gap has a
$d_{x^2-y^2}$ form when projected onto the Fermi surface, which has
nodes on the $E_{-}$ band Fermi surface sheets, but is approximately
constant in magnitude and changing sign between the two sheets on the
$E_{+}$ band Fermi surface sheets. As a result, the average of the gap
on the Fermi surface is vanishing, but even in the absence of the
spin-orbit coupling $\lambda$ the orbital structure of the bands gives
some protection against disorder. Each of the two sheets of the
$E_{+}$ band Fermi surface in this case are composed of states
belonging predominantly to one orbital or the other, due to the
dominant contribution from $\varepsilon_{z}$, and so any impurity
averaging of the gap on the Fermi surface requires some disorder with
a non-trivial orbital structure to rapidly suppress the
superconductivity.}

\subsubsection{$d$-wave gaps}

The $B_{1g}$ and $B_{2g}$ $s$-wave states have $d$-wave form-factors
when projected onto the states at the Fermi surface. We have
previously demonstrated~\cite{Cavanagh2020} that momentum-dependent pairing
states that belong to an irrep with an unconventional $s$-wave state
will generally inherit some of the robustness of that $s$-wave state
again disorder, due to overlap of the two gaps at the Fermi
surface. Solving the linearized gap equation for a multi-component gap, the 
generalization of Eq. \ref{eq:AG2band} for momentum-dependent gaps is
\begin{align}
  \log\left(\frac{T_c}{T_{c0}}\right) =& \sum_{j=\pm}\bar{R}_j\left[\psi\left(\frac{1}{2}\right) -   \psi\left(\frac{1}{2} + \frac{1}{4\pi k_BT_c\bar{\tau}_{\nu,j}}\right)\right]\nonumber\\
  & + \sum_{j=\pm}R_j\left[\psi\left(\frac{1}{2}\right) -   \psi\left(\frac{1}{2} + \frac{1}{4\pi k_BT_c\tau_{j}}\right)\right],\label{eq:modified_AG}
\end{align}
where the first line accounts for the overlap with the unconventional
$s$-wave states. 
The parameters $\bar{R}_j$ in this expression are
reduced compared to those of the purely $s$-wave case, as they satisfy 
$\sum_{j=\pm} R_j +\bar{R}_j=1$. 

Equation~\ref{eq:modified_AG} is distinguished by 
the presence of more than one effective scattering rate, which 
can significantly alter the shape of the disorder curve. This can be
understood more concretely by considering the limiting behavior of
Eq. \ref{eq:modified_AG} at both strong and weak disorder, where
  the curve is characterized by a single effective scattering rate. A Taylor expansion in the weak disorder limit for Eq. \ref{eq:modified_AG} gives $T_c\approx T_{c,0} -\pi/8k_B\tau_{\text{WD}}$,
dependent on a single average effective scattering rate
\begin{equation}
\tau_{\text{WD}}^{-1} = \sum_{i=\pm} \bar{R}_i\bar{\tau}_i^{-1}+R_i \tau_i^{-1},
\end{equation}
while for strong disorder the critical temperature can be shown to
vanish at a disorder strength given by the usual expression
\cite{Mineev1999, Mineev2007} $\log(2\pi k_B T_{c,0}
\tau_{\text{SD}})=-\psi(1/2)$ with an effective scattering rate
\begin{equation}
\tau_{\text{SD}}^{-1} = \prod_{i=\pm} \left(\bar{\tau}_i^{-1}\right)^{\bar{R}_i}\left(\tau_i^{-1}\right)^{R_i}\,.
\end{equation}
In general for a multi-band system
$\tau_{\text{WD}}\neq\tau_{\text{SD}}$, except when only a single
scattering rate is present, such as for an unconventional $s$-wave
state with a single band at the Fermi level.

\begin{figure}
	\begin{overpic}[width=\columnwidth]{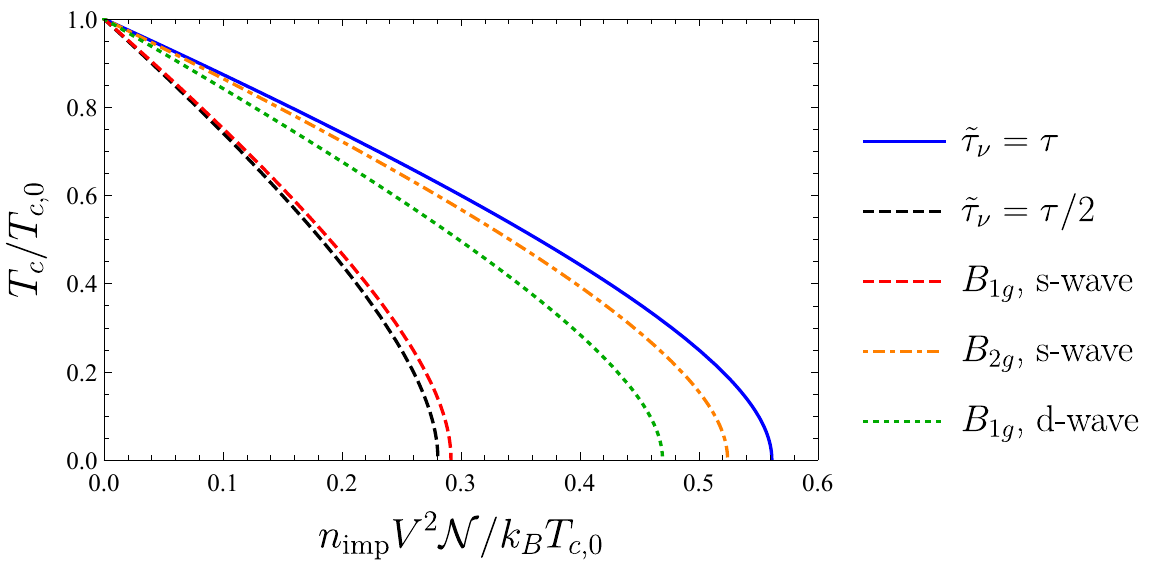}
	\end{overpic}
  \caption{Robustness against ``magnetic'' ($\lambda_{\alpha}=-1$)
    disorder for the $s$-wave $B_{1g}$  and $B_{2g}$ gaps, as well as
    the $d$-wave $B_{1g}$ gap, in the two-orbital model of the iron 
      pnictides, with spin-orbit coupling $\lambda=0.1\left|t_1\right|$.\label{fig:FeAs_Bg_TRSB}}
\end{figure}

\begin{figure}
	\begin{overpic}[width=\columnwidth]{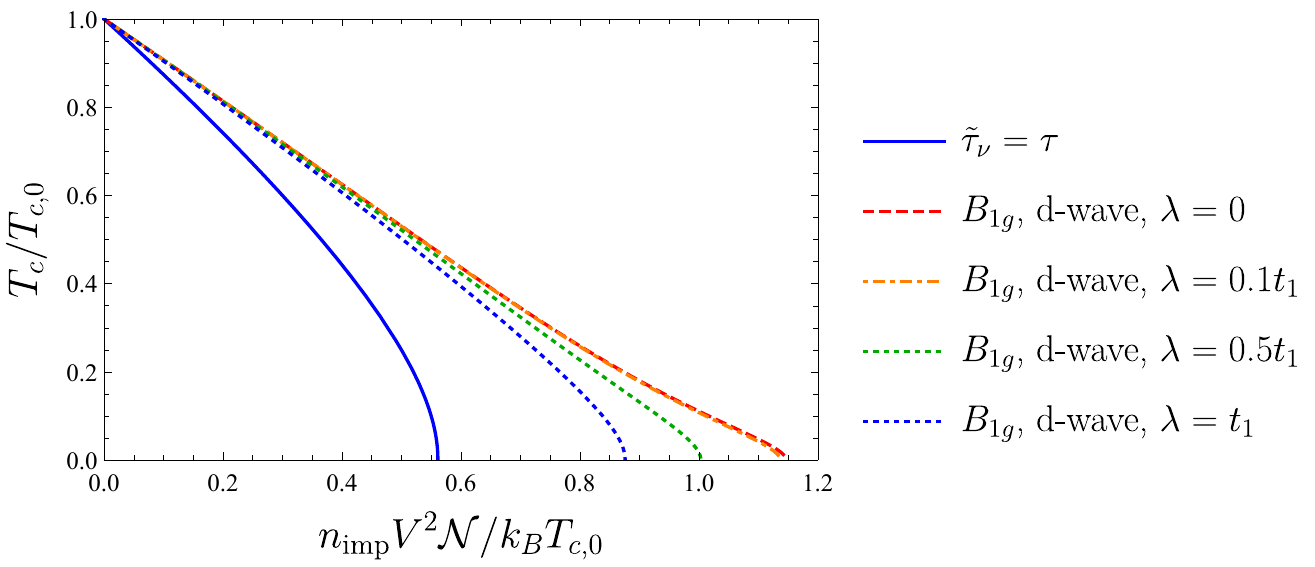}
	\end{overpic}
  \caption{Robustness against ``nonmagnetic'' ($\lambda_{\alpha}=+1$)
    disorder for  $B_{1g}$ $d$-wave gaps in the two-orbital model of the
    iron pnictides, for various values of the spin-orbit coupling $\lambda$. \label{fig:FeAs_BgDWaveCurves}}
\end{figure}

To illustrate, we consider an orbitally-trivial $d$-wave $B_{1g}$
state in the two-band iron pnictide model with $\Delta=\Delta_0 (\cos k_x-\cos k_y)\gamma^0U_T/2$. The
lowest-order contribution to the anomalous self-energy for this gap is proportional to the corresponding unconventional $s$-wave potential
\begin{align}
\Sigma^{(0)}_{2}  &=  \Delta_0\sum_\alpha\pi
                    n_{\text{imp},\alpha}|\tilde{V}_\alpha|^2\lambda^{(1)}_\alpha\notag\\
&\phantom{=}
                    \times\sum_{j=\pm}
                    \frac{j{\cal
                    N}_j}{4\left|\tilde{\omega}_{n,j}\right|} 
\langle (\cos k_x-\cos k_y)\varepsilon_z\rangle_{j}\gamma^1 U_T\,.
\end{align}
Since $\varepsilon_z$ also belongs to $B_{1g}$, the Fermi surface
average in this expression - the overlap between the $d$-wave and
$s$-wave states - is generally nonzero. We plot the critical
temperature against the ``nonmagnetic'' (with respect to the $s$-wave $B_{1g}$ state) disorder strength for this state in Fig.  \ref{fig:FeAs_BgDWaveCurves}. 
The $d$-wave state is considerably less
robust against ``nonmagnetic'' disorder than the $s$-wave state, 
  but still much more 
robust than predicted by the Abrikosov-Gor'kov result (solid line). The shape of
the curve also differs noticeably from the typical Abrikosov-Gor'kov
curve, in particular for weak spin-orbit coupling: while the
critical temperature is initially suppressed linearly with disorder,
the gradient decreases with increasing disorder, and clearly 
$\tau_{\text{WD}}^{-1}>\tau_{\text{SD}}^{-1}$. The robustness of the $d$-wave state against ``magnetic'' disorder, included in Fig. \ref{fig:FeAs_Bg_TRSB} is again influenced by the overlap with the unconventional $s$-wave state, which in this case reduces the robustness relative to a single-band $d$-wave gap.

\subsection{The $A_{1g}$ irrep}

As was the case for the Dirac system considered above, the behavior of
the two-channel $A_{1g}$ state is more complicated than that of the
single channel superconducting states. The general pairing
  potential in this state is given by
  \begin{equation}
\Delta =  \left[\Delta_0^{(0)}\gamma^{0}+ \Delta_{0}^{(5)}\gamma^{5}\right]U_T,\label{eq:Raghu_2comp_gap}
\end{equation}
where any combination of the two $s$-wave channels is allowed. Projected into the band-pseudospin basis, we have
\begin{equation}
\Delta \rightarrow \Delta_\pm = \left(\Delta_0^{(0)} \pm \hat{\lambda}\Delta_{0}^{(5)}\right)is_y
\end{equation}
This state is typically fully-gapped, and in the presence of
spin-orbit coupling it will change sign between the $+$ and $-$ bands 
when the ratio $\Delta_0^{(0)}/\Delta_{0}^{(5)}$ is sufficiently small.

In general we can have attractive interactions in both the
  conventional and unconventional channels, which we label $g_0$ and
  $g_5$, respectively. The critical temperature $T_c$ of the mixed
  pairing state has a complicated form
complicated expression, but for realistic
parameters we find that the actual $T_c$ is very close to the greater
of $T_{c,0}^{(0)}$ and $T_{c,0}^{(5)}$, which are the critical
temperatures of each channel in the absence of the other. These
have the explicit form 
\begin{align}
T_{c,0}^{(0)} = & \frac{2e^\gamma}{\pi}\Lambda \exp\left(-\frac{1}{g_0[{\cal N}_++{\cal N}_-]}\right)\\
T_{c,0}^{(5)} = & \frac{2e^\gamma}{\pi}\Lambda \exp\left(-\frac{1}{g_5[\langle\hat{\lambda}^2\rangle_+{\cal N}_++\langle\hat{\lambda}^2\rangle_-{\cal N}_-]}\right)
\end{align}
where $\Lambda$ is a cut-off and $\gamma$ is Euler's constant. 

The lowest-order  contribution to the anomalous self-energy is
\begin{widetext}
\begin{align}
\Sigma^{(0)}_{2}  &= \sum_\alpha\pi n_{\text{imp},\alpha}|\tilde{V}_\alpha|^2 \sum_{j=\pm} \frac{{\cal N}_j}{2\left|\tilde{\omega}_{n,j}\right|} \begin{pmatrix}
\Delta_0^{(0)} \\
\Delta_0^{(5)}
\end{pmatrix}^{\text{T}}
\begin{bmatrix}
\lambda^{(0)}_\alpha &   \lambda^{(5)}_\alpha j\langle \hat{\lambda}\rangle_{j} \\
\lambda^{(0)}_\alpha j\langle\hat{\lambda}\rangle_{j} & \lambda^{(5)}_\alpha\langle\hat{\lambda}^2\rangle_{j}
\end{bmatrix}
\begin{pmatrix}
\gamma^0 \\
\gamma^5
\end{pmatrix}U_T,\label{eq:Sig20_Raghu_A1g}
\end{align}
\end{widetext}
where we see that the disorder only couples the two channels if the
spin-orbit coupling is nonzero. This expression is similar to
  the self-energy for the $A_{1g}$ states in Cu$_x$Bi$_2$Se$_3$ Eq.~\ref{eq:Sig20_BiSe_A1g}, where the conventional and
unconventional $s$-wave $A_{1g}$ states are only coupled when the
mass $m$ is nonzero. 
In contrast with the case of Cu$_x$Bi$_2$Se$_3$, however, the
general $A_{1g}$ state is not completely robust against
``nonmagnetic'' disorder.  This is a direct consequence of the
presence of 
both bands at the Fermi energy and the changing sign of the
unconventional gap component between the two bands. 
The general expression for the critical temperature is lengthy
  and will not be presented here; numerical solutions of the
  linearized gap equation are plotted in
  Figs. \ref{fig:FeAs_A1gCurves} and \ref{fig:FeAs_A1g_LScale}. 

\subsubsection{Purely unconventional pairing}

In the limit that the pairing potential vanishes in the conventional $s$-wave channel, $g_0=0$, the superconducting gap is purely unconventional, with equal magnitude and opposite sign on the two Fermi surfaces. 
The critical temperature for the purely unconventional state is given by Eq. \ref{eq:AG2band}, with one of the two effective scattering rates vanishing in the presence of TRS disorder, $\bar{\tau}_{-}^{-1}=0$,
\begin{equation}
  \log\left(\frac{T_c}{T_{c0}}\right) = \bar{R}_{+}\left[\psi\left(\frac{1}{2}\right) -   \psi\left(\frac{1}{2} + \frac{1}{4\pi k_BT_c\bar{\tau}_{+}}\right)\right].
\end{equation}
Significantly, despite the absence of a contribution due to $\bar{\tau}_{-}$, the parameter $\bar{R}_{-}$, which characterizes the overlap of the unconventional $s$-wave state with the conventional, is non-vanishing, and as such $\bar{R}_{+} < 1$. Specifically, for TRS disorder ($\lambda_{\alpha}^{(0)}=+1$)
\begin{align}
\bar{R}_{+} &= \frac{\mathcal{N}_{+}\mathcal{N}_{-}\left(\langle\lambda\rangle_{+}+\langle\lambda\rangle_{-}\right)^2}{\left(\mathcal{N}_{+}+\mathcal{N}_{-}\right)\left(\mathcal{N}_{+}\langle\lambda^2\rangle_{+}+\mathcal{N}_{-}\langle\lambda^2\rangle_{-}\right)}\nonumber\\
\bar{\tau}_{+}^{-1}&=\frac{\pi\sum_{\alpha} n_{\text{imp},\alpha} |V_\alpha|^2}{2}\left(\mathcal{N}_{+}+\mathcal{N}_{-}\right)\left(1-\lambda_{\alpha}^{(5)}\langle\lambda\rangle_{+}\langle\lambda\rangle_{-}\right),
\end{align}
and the effective  scattering rate is simply the total interband scattering rate.

\begin{figure}
	\begin{overpic}[width=\columnwidth]{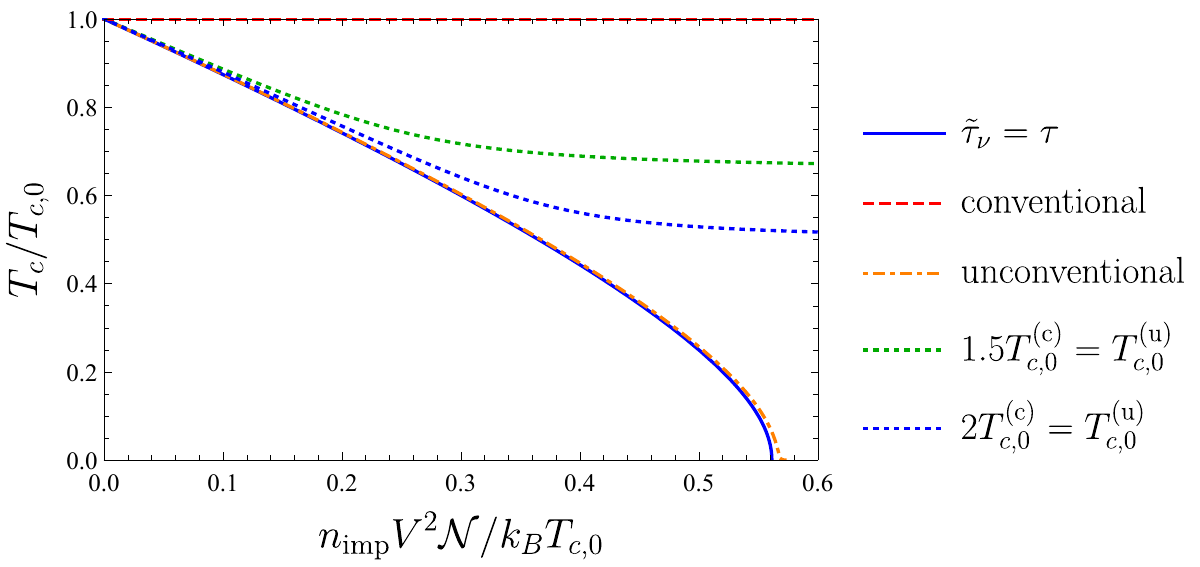}
	\put (30,40) {$\lambda^{(0)}_{\alpha}=\lambda^{(5)}_{\alpha}=+1$}
	\end{overpic}\\
	\begin{overpic}[width=\columnwidth]{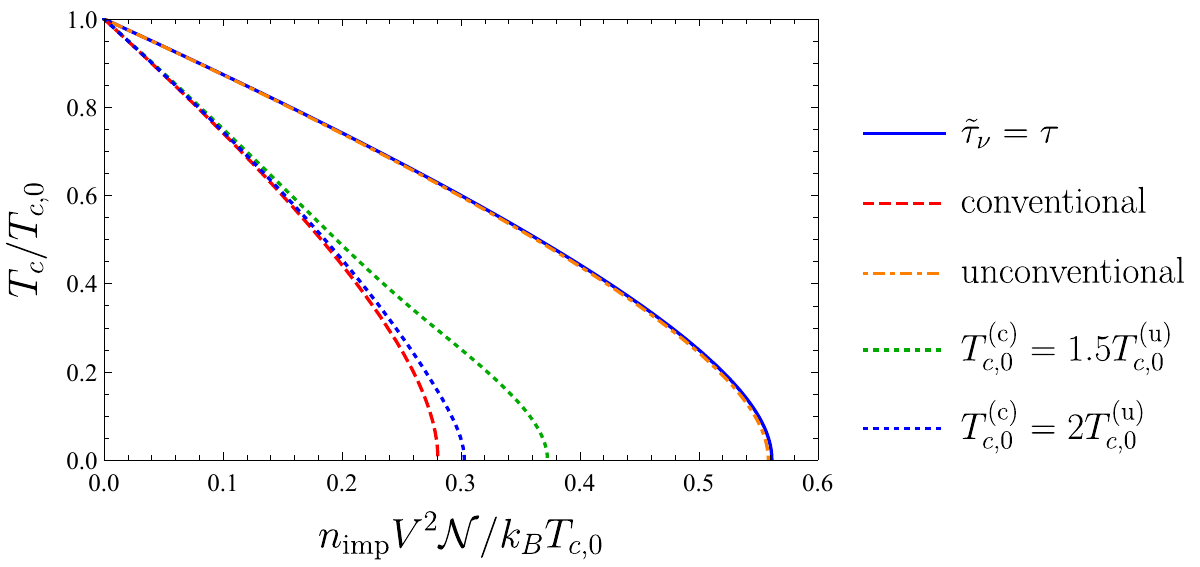}
	\put (30,40) {$\lambda^{(0)}_{\alpha}=\lambda^{(5)}_{\alpha}=-1$}
	\end{overpic}
  \caption{Robustness of the multi-channel $s$-wave $A_{1g}$ gap against against TRS (top) and TRSB (bottom) disorder, in the two-orbital model for iron oxypnictides with weak spin-orbit coupling $\lambda=0.1\left|t_1\right|$. $\lambda^{(0)}_{\alpha}$ is defined for the conventional $s$-wave state, and we find that the influence of $\lambda^{(5)}_{\alpha}$ for the unconventional gap is minimal. The states with pairing in both channels are labeled by the relative critical temperatures in the clean limit for the conventional ($T_{c,0}^{(\text{c})}$) and unconventional ($T_{c,0}^{(\text{u})}$) channels. \label{fig:FeAs_A1gCurves}}
\end{figure}

For weak disorder, the suppression of $T_c$ is linear in $\bar{\tau}_{+}^{-1}$, but in the strong disorder limit, we find
\begin{equation}
T_c \propto \tau_0^{\left(\frac{1-\bar{R}_{-}}{\bar{R}_{-}}\right)} \sim \left|V\right|^{2\left(\frac{\bar{R}_{-}-1}{\bar{R}_{-}}\right)}\label{eq:ExpTc}
\end{equation}
and $T_c$ is exponentially suppressed for $0<\bar{R}_{-}<1$. Interestingly, we find that even in the complete absence of conventional pairing the superconductivity retains some residual robustness against TRS disorder, due to the overlap with the conventional state. This effect is evident in the small exponential tail in the strong disorder limit for the purely unconventional curve in Fig. \ref{fig:FeAs_A1gCurves} close to $T_c=0$. 
For the particular model parameters adopted here we find that $\tilde{R}_{-}\ll 1$, 
and so for strong disorder $T_c$ is nearly indistinguishable from zero 
for the purely unconventional pairing state.

In contrast, for TRSB disorder, $\bar{\tau}_{\pm}^{-1}\neq 0$, and the superconductivity is completely  suppressed for strong disorder. In fact, the influence of the conventional $s$-wave state is slightly detrimental to the purely unconventional $s$-wave state (see Fig. \ref{fig:FeAs_A1gCurves}), due to the greater sensitivity of the conventional gap to TRSB disorder.

\subsubsection{Coexisting conventional and unconventional pairing}

\begin{figure}
	\begin{overpic}[width=\columnwidth]{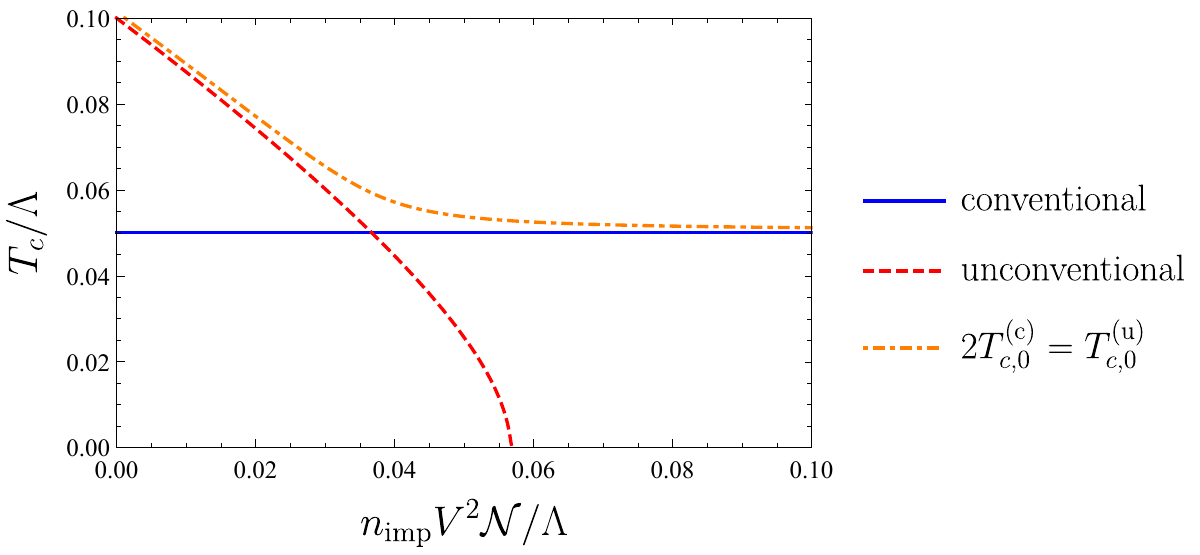}
	\put (28,38) {$\lambda^{(0)}_{\alpha}=\lambda^{(5)}_{\alpha}=+1$}
	\end{overpic}\\
	\begin{overpic}[width=\columnwidth]{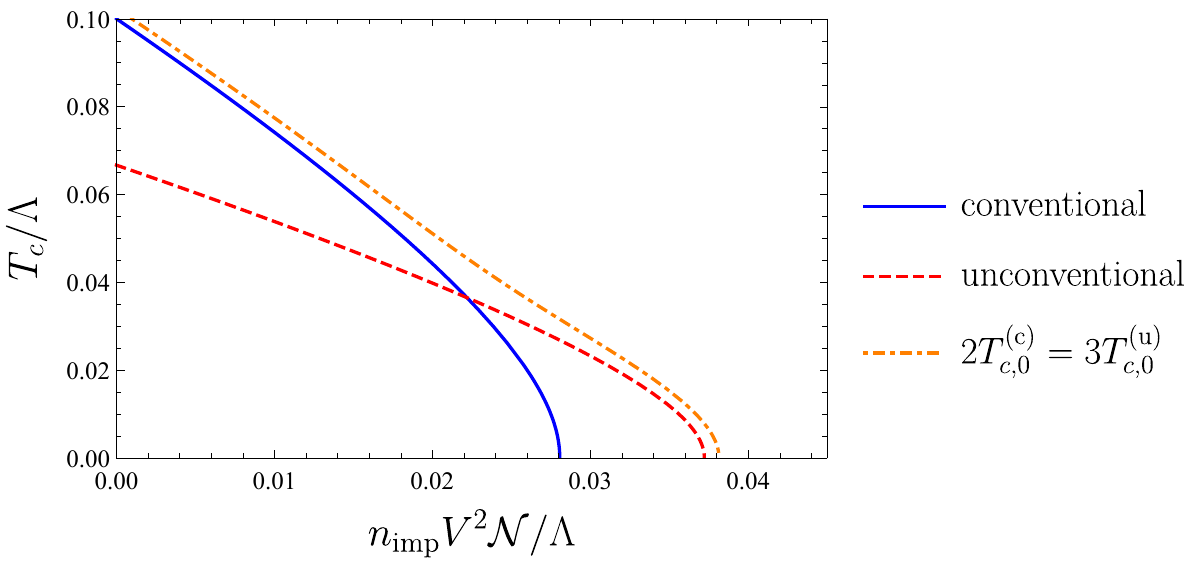}
	\put (28,38) {$\lambda^{(0)}_{\alpha}=\lambda^{(5)}_{\alpha}=-1$}
	\end{overpic}
  \caption{Robustness of the multi-channel $s$-wave $A_{1g}$ gap against against `nonmagnetic' ($\lambda^{(0)}_{\alpha}=\lambda^{(5)}_{\alpha}=+1$, top) and `magnetic' ($\lambda^{(0)}_{\alpha}=\lambda^{(5)}_{\alpha}=-1$, bottom) disorder, in the two-orbital model for iron oxypnictides with weak spin-orbit coupling $\lambda=0.1\left|t_1\right|$. Unlike Fig. \ref{fig:FeAs_A1gCurves}, both critical temperature and disorder strength are scaled by $\Lambda$, rather than $T_{c,0}$ (with $T_{c,0}^{(\text{c})}=0.1\Lambda$ and $T_{c,0}^{(\text{u})}=0.05\Lambda$).  \label{fig:FeAs_A1g_LScale}}
\end{figure}

In the presence of  time-reversal symmetry preserving
disorder (with $\lambda^{(0)}_{\alpha}=+1$ for the conventional gap
component), any general mixture of the two channels with $T_{c,0}^{(\text{5})}< T_{c,0}^{(\text{0})}$
is completely robust, whereas a state with 
$T_{c,0}^{(\text{5})}>T_{c,0}^{(\text{0})}$ is sensitive to
disorder. 
In the latter case, the critical temperature follows closely the
curve of the purely unconventional state at weak disorder, but saturates at the
critical temperature $T_{c,0}^{(0)}$ of the purely conventional state
in the strong-disorder limit. As seen in
Fig. \ref{fig:FeAs_A1g_LScale}, this crossover occurs when the
critical temperature for the purely-unconventional state falls below
that of the purely-conventional state, with the critical temperature
of the mixed state closely tracking the higher of the two. This can
hence be
interpreted as a disorder-induced crossover from a state where the unconventional pairing
dominates to a state where the conventional pairing is
dominant. This crossover between $s^{\pm}$-wave and conventional
$s$-wave states has been extensively studied in models where the
orbital degree of freedom is not explicitly included~\cite{Efremov2011,Efremov_2013,Stanev2014}.

For time-reversal symmetry breaking disorder potentials
  ($\lambda^{(0)}_{\alpha}=-1$) there may alternatively exist a crossover in the
intermediate disorder strength regime when
$T_{c,0}^{(5)}<T_{c,0}^{(0)}$. As
shown in Fig.~\ref{fig:FeAs_A1g_LScale}, this
crossover is less general than that for TRS disorder, since it requires
  that the unconventional state is sufficiently competitive with the
  conventional and has
  sufficiently small effective scattering rate, so that the termination
  point of the purely-unconventional state is at stronger disorder
  strength. If the unconventional state dominates in the clean limit
(i.e. $T_{c,0}^{(5)}>T_{c,0}^{(0)}$), the critical temperature closely
tracks the curve for purely-unconventional pairing, and no crossover
is observed. These contrasting
    crossover effects could be used to evidence the dominant component
    in the clean limit of a mixed pairing state.

\subsubsection{The momentum-dependent $s_{\pm}$-wave $A_{1g}$ gap}

Finally, we turn our attention to a momentum-dependent
$s_{\pm}$-wave singlet state,
\begin{equation}
\Delta = \Delta_0\cos(k_x)\cos(k_y) U_T\,.
\end{equation}
The form factor ensures that the gap on the $+$ and $-$ Fermi surfaces
has opposite sign, similar to the unconventional $s$-wave state. 
In this case, the robustness depends on the overlap of the
momentum-dependent state, with $f_{\bm{k}}=\cos(k_x)\cos(k_y)$, and
both $A_{1g}$ $s$-wave states, as well as the robustness of those
states and the effect of the normal state scattering rate.
 The lowest-order contribution to the anomalous self-energy is defined by the overlap of the $s_{\pm}$-wave state with the conventional and unconventional $s$-wave states,
\begin{widetext}
\begin{equation}
\Sigma^{(0)}_{2}  =\pi \sum_\alpha \frac{n_{\text{imp},\alpha}|\tilde{V}_\alpha|^2}{2} \sum_{j=\pm} \left[\lambda^{(0)}_\alpha\frac{{\cal N}_j\langle f_{\bm{k}}\rangle_{j}}{\left|\tilde{\omega}_{n,j}\right|}\gamma^0 + j\lambda^{(5)}_\alpha\frac{{\cal N}_j\langle f_{\bm{k}}\hat{\lambda}\rangle_{j}}{\left|\tilde{\omega}_{n,j}\right|}\gamma^5\right]U_T,
\end{equation}
\end{widetext}
and the overlap with the unconventional $s$-wave state depends on the spin-orbit coupling magnitude $\lambda$. Ultimately, the critical temperature is given by an expression of the form Eq. \ref{eq:modified_AG}, with one effective scattering rate vanishing $\bar{\tau}^{-1}_{-}=0$ for TRS disorder.

As in the case of the purely unconventional $s$-wave state, $T_c$ is exponentially suppressed, as described by Eq. \ref{eq:ExpTc}, for strong TRS disorder, but the much greater overlap of the $s_{\pm}$-wave gap with the conventional $s$-wave enhances this effect. 

Spin-orbital effects play a significant role in determining the robustness of the $s_{\pm}$-wave state. Increasing the spin-orbit coupling $\lambda$ increases both the interband scattering and, more significantly, the overlap with the unconventional $s$-wave state. Unlike the momentum-dependent $B_{1g}$ $d$-wave state, the existence of the unconventional $s$-wave state is detrimental to the robustness of the momentum dependent $A_{1g}$ $s_{\pm}$-wave gap against disorder, and increasing the overlap between the states reduces the overall robustness. For TRSB disorder, the overlap with the conventional $s$-wave state is detrimental, and increasing the overlap with the unconventional $s$-wave state by increasing the spin-orbit coupling increases the robustness.

\begin{figure}
	\begin{overpic}[width=\columnwidth]{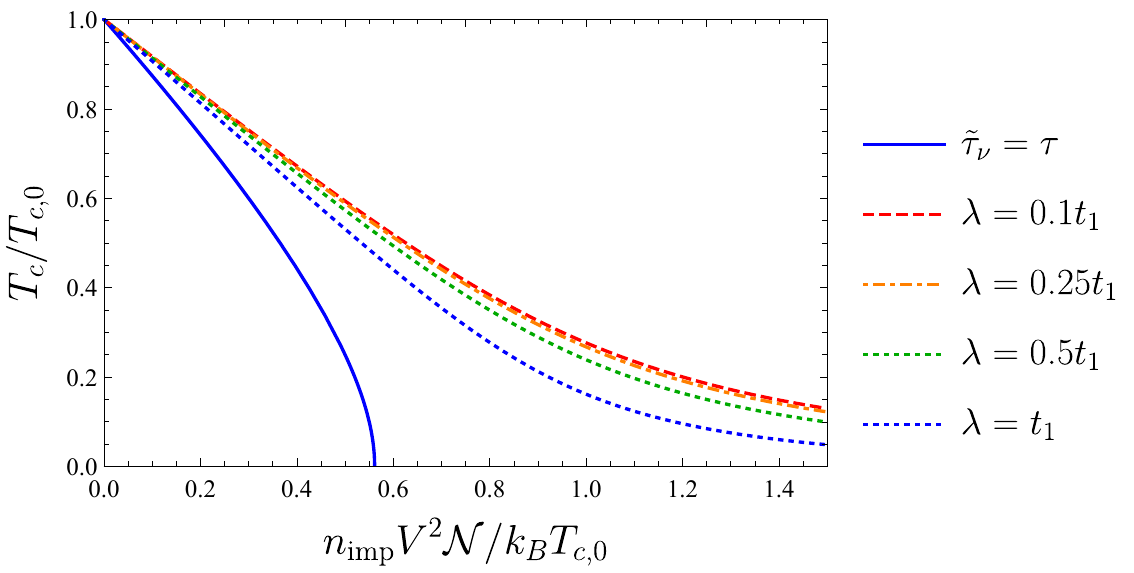}
	\put (30,40) {$\lambda^{(0)}_{\alpha}=\lambda^{(5)}_{\alpha}=+1$}
	\end{overpic}\\
	\begin{overpic}[width=\columnwidth]{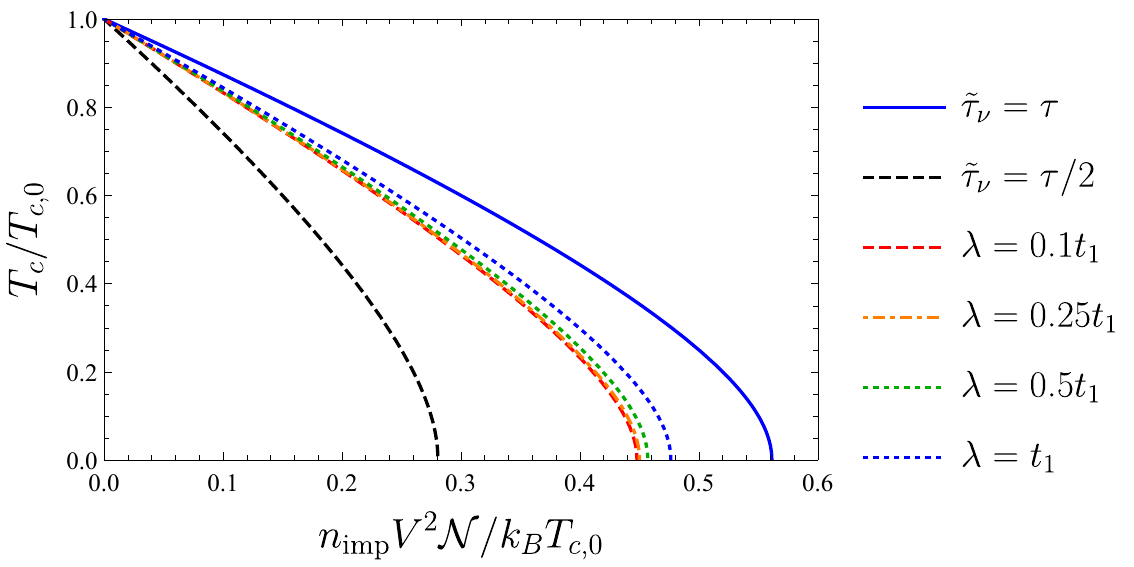}
	\put (30,40) {$\lambda^{(0)}_{\alpha}=-1$, $\lambda^{(5)}_{\alpha}=+1$}
	\end{overpic}
  \caption{Robustness of the momentum-dependent $s_{\pm}$-wave $A_{1g}$ gap against against TRS (top) and TRSB (bottom) disorder, in the two-orbital model for iron oxypnictides with various spin-orbit coupling strengths.  The influence of $\lambda^{(5)}_{\alpha}$, the factor for the unconventional $s$-wave state has only a minor influence on the robustness and so we consider only $\lambda^{(5)}_{\alpha}=+1$ here. \label{fig:FeAs_A1g_sPM}}
\end{figure}

The results of this calculation are presented in Fig. \ref{fig:FeAs_A1g_sPM}, for TRS and TRSB disorder, under the assumption that there is pairing only in the momentum-dependent channel.

\section{Discussion}\label{Sec:Discussion}

The existence of unconventional $s$-wave pairing states in systems with additional internal degrees of freedom has significant consequences  for the robustness of superconductivity against disorder. The general framework we have presented provides a straightforward, analytically tractable method to predict the robustness of a particular superconducting state in a given system, and also easily accounts for known results in systems of interest \cite{MichaeliFu2012, Dentelski2020, ScheurerTimmons2020, Cavanagh2020}. The superconducting fitness is of critical importance in determining the robustness, as evidenced clearly in the robustness of the $B_{1g}$ gap in the iron-pnictide model (see Fig. \ref{fig:FeAs_BgCurves}), which has a nearly perfect fitness on one band at the Fermi surface. 

Interestingly, the fitness of the superconducting gap with respect to
the disorder potential plays perhaps an even more significant role. In
our framework, the parameter $\lambda_{\alpha}=\pm 1$ encapsulates the
fitness with respect to disorder, and determines whether the
spin-orbital texture acts to enhance (if $\lambda_\alpha=+1$, which
requires
$\tilde{V}_\alpha\tilde{\Delta}_{\nu}-\tilde{\Delta}_{\nu}\tilde{V}_\alpha^T=0$)
or reduce (if $\lambda_\alpha=-1$,
$\tilde{V}_\alpha\tilde{\Delta}_{\nu}-\tilde{\Delta}_{\nu}\tilde{V}_\alpha^T\neq
0$) the robustness. 
This is consistent with Anderson's theorem: for the conventional
$s$-wave singlet state of single-band systems,  TRS disorder potentials are perfectly fit and TRSB potentials perfectly unfit. 
As we have demonstrated, when additional degrees of freedom are
present, there will exist TRSB disorder potentials for which a given
unconventional $s$-wave state is fit, and as a result the
superconductor will be more robust against certain forms of TRSB disorder than TRS disorder potentials. The fitness with respect to the disorder potential is therefore a more versatile definition when considering the influence of disorder, as opposed to whether the disorder preserves or breaks time-reversal symmetry.

The important role played by the superconducting fitness in 
determining the robustness against scalar disorder can be understood 
via a canonical transformation, $c_{\bm k} \rightarrow 
\exp(i\frac{\pi}{4}\tilde{\Delta}_\nu U_T^\dagger)c_{\bm k}$, that 
maps the unconventional $s$-wave state to a conventional $s$-wave 
state \cite{FuBerg2010,Cavanagh2020,Dentelski2020}. Under such a 
transformation, components of the Hamiltonian for which the gap is perfectly fit
(i.e. $\varepsilon_{\bm{k},i}\gamma^i\tilde{\Delta}_{\nu}-\tilde{\Delta}_{\nu}[\varepsilon_{\bm{k},i}\gamma^i]^T=0$)
are mapped onto TRS terms, whereas the unfit components are
transformed into TRSB terms. The superconductivity is therefore robust
against the former terms, but is destabilized by the
latter. Similarly, this
transformation also maps disorder potentials with $\lambda_{\alpha}=1$
onto TRS disorder potentials, whereas the  disorder potentials with
$\lambda_{\alpha}=-1$ are mapped onto TRSB potentials. 
The pairing consequently has 
enhanced robustness against disorder potentials with 
$\lambda_{\alpha}=+1$, but the mapping of the unfit elements of
  the Hamiltonian to TRSB terms violates Anderson's theorem, 
preventing perfect robustness. On the other hand, the unfit terms of the 
Hamiltonian reduce the magnitude of the effective scattering rate 
due to the disorder potentials with $\lambda_{\alpha}=-1$ 
relative to the value for a conventional $s$-wave, and thereby
  lead to an enhanced robustness.

We have additionally demonstrated the significant role played by the number of bands that cross the Fermi level. As a clear example, consider the unconventional $s$-wave $A_{1g}$ pairing state in the two models we consider: in the Dirac system a single band crosses the Fermi surface and the unconventional state is indistinguishable from the conventional $s$-wave, while for the iron pnictide model the unconventional gap component changes sign between the two bands' Fermi surfaces. In the first case, the general $A_{1g}$ pairing state is always completely robust against TRS disorder, while in the second the unconventional state is sensitive (to a degree determined by the superconducting fitness) to all disorder as is the unconventional component of a general pairing state.

\subsection{Relationship to other work}

As we have noted previously, our framework readily accounts for, and significantly generalizes, recent results for Dirac superconductors in the presence of trivial and non-trivial disorder potentials \cite{Dentelski2020, MichaeliFu2012}. More generally, our results are consistent with recent proposals for generalizations of Anderson's Theorem \cite{Scheurer2015, ScheurerTimmons2020}. The generalized Anderson's theorem proposed in Ref. \cite{ScheurerTimmons2020} is of particular interest, being completely consistent with our own framework, and demonstrating that our results can be straightforwardly generalized to systems with more than two bands. 
A key point of difference between our result and the generalized theorem of Ref. \cite{ScheurerTimmons2020}, is our explicit treatment of the spin and orbital degrees of freedom.

Briefly, we wish to compare our results for Dirac systems with recent findings which are apparently inconsistent with our results. Sato and Asano \cite{Sato2020} found an even-orbital-parity spin-singlet $s$-wave state in Cu$_x$Bi$_2$Se$_3$ (belonging to the $A_{2u}$ irrep) to be robust against disorder. In the weak disorder limit our results are consistent, but for strong disorder a gradual exponential suppression is seen in Ref. \cite{Sato2020}, while our framework predicts that superconductivity is completely suppressed for sufficiently strong disorder. 

In calculating the anomalous part of the self-energy, we have made use of the self-consistent Born approximation, while the authors in \cite{Sato2020} consider only the lowest-order Born approximation. 
While the distinction between the two is not significant in single-band materials, when considering multi-orbital superconductors the lowest-order approximation fails in the strong disorder limit. The result of Ref. \cite{Sato2020} is, predictably, consistent with our own for weak disorder, but the two diverge for larger disorder strengths, with Sato and Asano predicting an exponential suppression of $T_c$ whereas we find a vanishing critical temperature for strong disorder.   

Andersen \emph{et al.} \cite{Ramires2019} recently demonstrated a complete robustness of superconductivity against a fit disorder potential in a Bi$_2$Se$_3$-based superconductor. As we have noted previously \cite{Cavanagh2020}, however, this complete robustness additionally relies on an implicit assumption that the superconducting state is perfectly fit.  Including this additional assumption brings their conclusion into agreement with our own framework as well as other recent results \cite{MichaeliFu2012, ScheurerTimmons2020}.

\section{Conclusions}\label{Sec:Conclusion}

We have presented a general framework, based on the self-consistent Born-approximation, to consider the robustness of superconductivity against various forms of disorder in systems with additional internal degrees of freedom and have highlighted the important role played by the superconducting fitness. Disorder potentials, $\tilde{V}_{\alpha}$, can generally be classified by the fitness of the superconducting state with respect to the disorder potential.  The superconducting fitness with regard  to the normal state Hamiltonian then defines the degree to which the state is robust against disorder. The spin-orbital texture, as encapsulated by the superconducting fitness, acts to enhance the robustness  against disorder for which the gap is fit, but reduces the robustness against unfit forms of disorder. We have also demonstrated how the robustness is influenced by the presence of multiple Fermi surfaces, where the presence of multiple effective scattering rate can significantly alter the  robustness, most noticeably for orbitally-trivial unconventional gaps.

\begin{acknowledgments}
  
The authors are thankful to David Dentelski, Jonathan Ruhman, Takumi Sato, Yasuhiro Asano, and Peter Orth for helpful discussions. This work was supported by the Marsden Fund Council from Government
funding, managed by Royal Society Te Ap\={a}rangi.


\end{acknowledgments}
%
%
%

\maketitle
\bibliography{references}

\end{document}